\documentclass[runningheads]{llncs}

\usepackage{amssymb,amsmath,amsfonts}
 \usepackage{latexsym}
 \usepackage[normalem]{ulem}
\usepackage{color}
 \usepackage{hyperref}
 \usepackage{tikz}
 \usepackage{appendix}
 \usepackage{algorithm}
 \usepackage[noend]{algpseudocode}

 \newtheorem{thrm}{Theorem}
 \newtheorem{lemm}[thrm]{Lemma}

 \newtheorem{obs}[thrm]{Observation}
 
 \newtheorem{defn}[thrm]{Definition}
 \newtheorem{examp}[thrm]{Example}

\AtBeginDocument{\renewcommand{\Omega}{\varOmega}}

 \def\s#1{\mbox{\boldmath $#1$}}
 \def\+{\!+\!}
 \def\itbf#1{\textit{\textbf{#1}}}

\def\s#1{\mbox{\boldmath $#1$}}

\def\+{\!+\!}
\def\-{\!-\!}

\def\itbf#1{\textit{\textbf{#1}}}

 \def\LF{\mbox{LF}}
 \def\EL{\mathcal L}

\begin{document}
 \title{$V$-Words, Lyndon Words and Galois Words
 \thanks{A preliminary version of this paper was presented at COCOA 2023. The main new contributions in this extended version are 
 Section~\ref{subsec_Galois} and 
 Section~\ref{sec_UMFF2cUMFF}.}}
\authorrunning{Daykin et al.}

 \author{Jacqueline W.\ Daykin\inst{1,2,3} \and
Neerja Mhaskar\inst{4}\thanks{Corresponding author.} \and
W.~F.~Smyth\inst{4}}

\institute{Department of Computer Science, Aberystwyth University, Wales\\
\email{jwd6@aber.ac.uk} \and
Department of Computer Science, Stellenbosch University, South Africa \and
Univ Rouen Normandie, INSA Rouen Normandie, Université Le Havre Normandie, Normandie Univ, LITIS UR 4108, F-76000 Rouen, France
\and
Department of Computing and Software, McMaster University, Canada
\email{\{pophlin,smyth\}@mcmaster.ca}}

\maketitle              
\begin{abstract}
We say that a family $\mathcal{W}$ of strings over $\Sigma^+$ forms a Unique Maximal Factorization Family (UMFF) if and only if every $\s{w} \in \mathcal{W}$ has a unique maximal factorization. Further, an UMFF $\mathcal{W}$ is called a circ-UMFF whenever it contains exactly one rotation of every primitive string $\s{x} \in \Sigma^+$.
$V$-order is a non-lexicographical total ordering on strings that determines a circ-UMFF. In this paper we propose a generalization of circ-UMFF called the substring circ-UMFF and extend combinatorial research on $V$-order by investigating connections to Lyndon words. Then we extend these concepts to any total order. Applications of this research arise in efficient text indexing, compression, and search problems.
\end{abstract}

\section{Introduction}\label{sect-intro}

$V$-order (Definition~\ref{def-V_order}) is a non-lexicographic global order on strings that was introduced more than a quarter-century ago \cite{DaD96,DD03}. Similar to conventional lexicographical order (\itbf{lexorder}), $V$-order string comparison can be performed using a simple linear time, constant space algorithm~\cite{ADRS14,ADRS15,ADRS14FI,DDS11,DDS13}, further improved in~\cite{ADKRS16,ADM18}. Much theoretical research has been done on this ordering~\cite{DDS11,DDS13,DS14}, including efficient construction of the so-called $V$-BWT or $V$-transform~\cite{DS14}, a variant of the lexicographic Burrows-Wheeler transform (BWT). 

In this paper, we further extend combinatorial research on $V$-order and circ-UMFFs. We first show that there are infinitely more $V$-words (Definition~\ref{Vword}) than Lyndon words (Definition~\ref{def-lyndon}). Then we study instances of circ-UMFFs having similar properties to $V$-words and/or Lyndon words. Finally, we propose a generalization of the circ-UMFF (Definition~\ref{def-subcirc}) called the substring circ-UMFF (Definition~\ref{def-subcircUMFF}) and show that for a generalized order $\mathcal{T}$, with order relation $\ll$, classes of border-free words exist that form 
circ-UMFFs and substring circ-UMFFs, respectively.

A useful tool in this work is a generalization of lexorder, which is based on letter comparisons, to lex-extension order, based on substring comparisons, first introduced in \cite{DD03}. This in turn allows generalizing Lyndon words to Hybrid Lyndon words \cite{DDS13}, which we apply here.

Using the framework of UMFFs, we explore new properties of Galois words. These words are defined analogously to Lyndon words except that, rather than lexorder, alternating lexicographic order (alternating lexorder) is applied. Starting with the relation $<$, the alternating variant processes letter comparisons from left to right alternating between the relations $<$ and $>$. While Lyndon words are necessarily border-free, the Galois variant defines both border-free and bordered words. Galois words are studied combinatorially in \cite{Reutenauer2005} and subsequently from an algorithmic perspective including practical applications in \cite{DBLP:journals/tcs/GiancarloMRRS20,HKYS24}.

All words in a circ-UMFF are border-free; thus in the general case Galois words do not form a circ-UMFF; indeed, examples show non-unique factorizations of strings into Galois words. Nevertheless, we show that the subset of {\it border-free} Galois words forms an UMFF.  Further, we characterize the structure of binary border-free Galois words in terms of a Hybrid Lyndon factorization.   We derive a Galois equivalent of the fundamental Lyndon result that the ordered concatenation of Lyndon words forms a Lyndon word --- the Galois version requires the concatenated word to be primitive.

In \cite{DD08} it is established that any binary border-free UMFF can be enlarged to a binary circ-UMFF. In view of this result, we propose here a constructive method in the finite case for generating a binary circ-UMFF from a border-free binary UMFF. We illustrate the algorithm by constructing an example of a circ-UMFF which consists of strings based on more than one method of string ordering --- a discovery we believe worthy of further investigation.


\section{Preliminaries}\label{sect-prelim}

A \itbf{string} (or \itbf{word}) is an array of elements drawn from a finite totally ordered set $\Sigma$ of cardinality $\sigma = |\Sigma|$, called the \itbf{alphabet}. The elements of $\Sigma$ are referred to as \itbf{characters} (\itbf{letters}).  We refer to strings using mathbold: \s{x}, \s{w} instead of $x,w$.  The \itbf{length} of a string $\s{w}[1..n]$ is $|\s{w}|=n$. The \itbf{empty string}
of length zero is denoted by $\s{\varepsilon}$. The set of all nonempty strings over the alphabet $\Sigma$ is denoted by $\Sigma^+$, with $\Sigma^* = \Sigma^+ \cup \s{\varepsilon}$. 
If $\s{x} = \s{u}\s{w}\s{v}$ for (possibly empty) strings $\s{u},\s{w},\s{v} \in \Sigma^{\ast}$,
then \s{u} is a \itbf{prefix}, \s{w} a \itbf{substring} or \itbf{factor},
and \s{v} a \itbf{suffix} of \s{x}. A substring \s{u} of \s{w} is said to be \itbf{proper} if $|\s{u}| < |\s{w}|$. 
A string \s{w} has a \itbf{border} \s{u} if \s{u} is both a proper prefix and a proper suffix of \s{w}. 
If \s{w} has only the empty border \itbf{$\varepsilon$}, then it is said to be \itbf{border-free}. 

For $\s{x} = \s{x}[1..n]$
and an integer sequence $0 < i_1 < i_2 < \cdots < i_k \le n$,
the string $\s{y} = \s{x}[i_1]\s{x}[i_2] \cdots \s{x}[i_k]$ is said to be a
\itbf{subsequence} of \s{x}, \itbf{proper} if $|\s{y}| < n$.
If $\s{x} = \s{u}^k$ (a concatenation of $k$ copies of \s{u})
for some nonempty string \s{u} and some integer $k > 1$,
then \s{x} is said to be a \itbf{repetition};
otherwise, \s{x} is \itbf{primitive}. We say $\s{x}$ has \itbf{period} $p$ if and only if for every $i \in 1..n-p$,
$\s{x}[i] = \s{x}[i+p]$; the shortest period of \s{x} is called \itbf{the period}.
A string $\s{y}=R_i(\s{x})$ is the $i^{\mbox{th}}$ \itbf{conjugate} (or \itbf{rotation}) of $\s{x}=\s{x}[1..n]$ if 
$\s{y} = \s{x}[i+1..n]\s{x}[1..i]$ for some $0 \leq i < n$ (so that $R_0(\s{x}) =\s{x}$).
The \itbf{conjugacy class} of \s{x} is the set $R_i(\s{x})$, $0 \leq i < n$, of all conjugates.

{In our examples, we often suppose that $\Sigma = \{a,b,c,\dots,z\}$, the Roman alphabet in its natural order, or $\Sigma = \{1,2,3,\dots,k\}$, the bounded natural numbers. The ordering of $\Sigma$ imposes \itbf{lexicographic order (lexorder)} on $\Sigma^+$.
However, in the case that $\s{x} = \s{u}^k$, a repetition, lexorder does not provide a unique ordering of the rotations of \s{x}: rotations $R_k, R_{2k},\ldots, R_{|\s{x}|}$ are all equal. 
To avoid this, we can append to each \s{x} a unique least symbol $\$$, so that all the rotations of \s{x}$\$$ are distinct and thus ordered, while the ordering of any pair \s{x}$\$$, \s{y}$\$$ is unaffected.
The rotations of \s{x}, sorted in ascending lexicographic order, form the \itbf{Burrows-Wheeler matrix}, so that, for
$\s{x} = abab\$$, we get the unique ordering
$$\begin{array}{ccccc}
\$ & a & b & a & b \\
a & b & \$ & a & b \\ 
a & b & a & b & \$ \\ 
b & \$ & a & b & a \\ 
b & a & b & \$ & a
\end{array}
$$
of which the last column, $bb\$aa$, is the \itbf{Burrows-Wheeler transform} (BWT) of \s{x}, from which the original string can be recovered in linear time \cite{BW94}. Observe the transformation of the input in the example, $abab\$ $ to $ b^2 \$ a^2 $, and this data clustering property is often exhibited with the BWT, and hence its use as a preprocessor or booster for the performance of memoryless compressors. While the BWT was originally introduced in the context of lossless text compression, subsequent applications span image compression, shape analysis in computer vision, efficient self-indexed compressed data structures, and its pattern matching properties are invaluable in bioinformatics with its highly repetitive data for tasks such as sequence alignment \cite{ABM08}.
}

{We now define another fundamental concept in this paper, the Lyndon word, which has deep connections with the theory of free Lie algebras and combinatorics on words:}

\begin{defn}
\label{def-lyndon}
A {\bf Lyndon word} \cite{CFL58} is a primitive string
that is minimum in lexorder $<$ over its conjugacy class.
\end{defn}
The following Lyndon factorization ({\itshape LF}) theorem is fundamental in stringology
and underpins the wide-ranging applications of Lyndon words,
{which include specialized string sorting tasks, digital geometry, musicology, the Burrows-Wheeler transform and data compression techniques:}
\begin{thrm}
\label{lLyndon-thrm}
\cite{CFL58} Any nonempty string $\s{x}$ can be written uniquely as a product
$\LF_{\s{x}} = \s{x} = \s{u}_1 \s{u}_2 \cdots \s{u}_k$ of $k \ge 1$ Lyndon words,
with $(\s{u}_1 \ge \s{u}_2 \ge \cdots \ge \s{u}_k)$.
\end{thrm}

For further stringological definitions and theory, see \cite{CHL2007,S03}.

\section{$V$-order}\label{sect-vorder}

In this section we start by defining $V$-order and describing some of its important properties used later in the paper.

Let $\s{x}=x_1x_2\cdots x_n$ be a
string over $\Sigma$. Define $h \in \{1,\ldots,n\}$ by $h = 1$ if $x_1 \le x_2 \le \cdots
 \le x_n$; otherwise, by the unique value such that $x_{h-1}>x_h \le x_{h+1} \le x_{h+2}
\le \cdots \le x_n$. Let $\s{x}^*=x_1x_2\cdots
x_{h-1}x_{h+1}\cdots x_n$, where
the star * indicates deletion of $x_h$. Write $\s{x}^{s*} =
(...(\s{x}^*)^{*}...)^{*}$ with $s \ge 0$ stars.
Let $g = \max\{x_1,x_2, \ldots ,x_n\}$,
and let $k$ be the number of occurrences of
$g$ in $\s{x}$.
Then the sequence $\s{x},\s{x}^*,\s{x}^{2*}, ...$
ends $g^{k},...,g^1,g^0=\s{\varepsilon}$.
From all strings \s{x} over $\Sigma$ we form the {\it star tree} (see {Example~\ref{ex-star_gen}}), where each string
$\s{x}$ labels a vertex and there is a directed edge upward from $\s{x}$
to $\s{x}^*$, with the empty string $\s{\varepsilon}$ as the root.

\begin{defn}[\bf {$V$-order}~\cite{DaD96}]
\label{def-V_order}
We define {\rm $V$-order} $\prec$ for distinct strings \s{x}, \s{y}. 
First $\s{x} \prec \s{y}$ if in the star tree \s{x} is in the path
$\s{y},\s{y}^*,\s{y}^{2*}, \ldots ,\s{\varepsilon}$. If $\s{x},\s{y}$ are not in a path, there exist
smallest $s,t$ such that $\s{x}^{(s+1)*}=\s{y}^{(t+1)*}$.  Let $\s{s}=\s{x}^{s*}$ and
$\s{t}=\s{y}^{t*}$; then $\s{s} \neq \s{t}$ but $|\s{s}| = |\s{t}| = m$ say.
Let $j \in [1..m]$ be the greatest integer such that $\s{s}[j] \ne \s{t}[j]$.
If $\s{s}[j]<\s{t}[j]$ in $\Sigma$ then $\s{x} \prec \s{y}$;
otherwise, $\s{y} \prec \s{x}$.
Clearly $\prec$ is a total
order on all strings in $\Sigma^{\ast}$.
\end{defn}


See the star tree path and star tree examples in 
Figures~\ref{fig-ex-Vorder-path} and~\ref{fig-ex-Vorder-gen}, respectively.

\begin{examp}
\label{ex-star_path}
[Star tree path] Figure~\ref{fig-ex-Vorder-path} illustrates the star tree for the case $\s{x} \prec \s{y}$ if in the star tree \s{x} is in the path
$\s{y},\s{y}^*,\s{y}^{2*}, \ldots ,\s{\varepsilon}$.  Consider the
$V$-order comparison of the strings $\s{x} = 929$ and $\s{y} = 922911$. The subscript $h$ indicates the $V$ letter to be deleted (defined above as $x_{h-1}>x_h \le x_{h+1} \le x_{h+2} \le \cdots \le x_n$). Since $929$ is in the path of star deletions of $922911$, therefore $929 \prec 922911$.
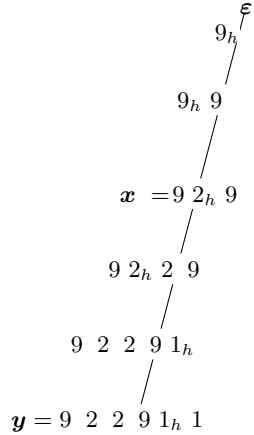
\begin{figure}[h]
\centering
\begin{tikzpicture}[x=-1cm, every node/.style={inner sep=0,outer sep=1mm, anchor=base, yshift=-3, fill=white}]
\newlength{\letterspace}
\setlength{\letterspace}{3.5mm}
\draw (5.75, 0) -- (7.2, -5.5);
\path[draw=white, line width = 3ex] (8.5, -5.5) node {$\s{y} =~$} 
-- ++(\letterspace, 0) node {$9$}
-- ++(\letterspace, 0) node {$2$} 
-- ++(\letterspace, 0) node {$2$}
-- ++(\letterspace, 0) node {$9$} 
-- ++(\letterspace, 0) node {$1_h$}
-- ++(\letterspace, 0) node {$1$};
\path[draw=white, line width = 3ex] (8, -4.5) node {$9$} 
-- ++(\letterspace, 0) node {$2$} 
-- ++(\letterspace, 0) node {$2$}
-- ++(\letterspace, 0) node {$9$} 
-- ++(\letterspace, 0) node {$1_h$};
\path[draw=white, line width = 3ex] (7.5, -3.5) node {$9$} 
-- ++(\letterspace, 0) node {$2_h$} 
-- ++(\letterspace, 0) node {$2$}
-- ++(\letterspace, 0) node {$9$};
\path[draw=white, line width = 3ex] (7, -2.5) node {$\s{x}~=~$} 
-- ++(\letterspace, 0) node {$9$}
-- ++(\letterspace, 0) node {$2_h$}
-- ++(\letterspace, 0) node {$9$};
\path[draw=white, line width = 3ex] (6.85, -1.25) 
-- ++(\letterspace, 0) node {$9_h$}
-- ++(\letterspace, 0) node {$9$}
 -- ++(\letterspace, 0);
 \path[draw=white, line width = 3ex] (6, -0.35) node {$9_h$};
 \path[draw=white, line width = 3ex] (5.75, 0) node {$\s{\varepsilon}$};
\end{tikzpicture}
\caption{\label{fig-ex-Vorder-path}
$929 \prec 922911$}
\end{figure}
\end{examp}

\begin{examp}
\label{ex-star_gen}
[Star tree] Figure~\ref{fig-ex-Vorder-gen} illustrates the star tree for the non-path case using the $V$-order comparison of the words $\s{x} = unique$ and $\s{y} = equitant$. As in the previous example, the subscript $h$ indicates the $V$ letter to be deleted (defined above as $x_{h-1}>x_h \le x_{h+1} \le x_{h+2} \le \cdots \le x_n$). The circled letters are those compared in alphabetic order (defined above as $\s{s}[j] \ne \s{t}[j]$). 
\begin{figure}[h]
\centering
\begin{tikzpicture}[x=-1cm, every node/.style={inner sep=0,outer sep=1mm, anchor=base, yshift=-3, fill=white}]
\setlength{\letterspace}{4mm}
 \path[draw=white, line width = 3ex] (5.75, 0) node {$u$};
 code for >
 \path (5.7, -1.75) node {$>$};
 \draw (5.5, -0.5) -- (1.5, -7.5);
 \draw (6, -0.5) -- (7.2, -5.5);
\path[draw=white, line width = 3ex] (2.6, -7.5) node {$e$} 
 -- ++(\letterspace, 0) node {$q$} 
 -- ++(\letterspace, 0) node {$u$} 
 -- ++(\letterspace, 0) node {$i$}
 -- ++(\letterspace, 0) node {$t$} 
 -- ++(\letterspace, 0) node {$a_h$} 
 -- ++(\letterspace, 0) node {$n$}
 -- ++(\letterspace, 0) node {$t$}; 
 \path[draw=white, line width = 3ex] (3, -6.5) node {$e$}
 -- ++(\letterspace, 0) node {$q$} 
 -- ++(\letterspace, 0) node {$u$} 
 -- ++(\letterspace, 0) node {$i$}
 -- ++(\letterspace, 0) node {$t$} 
 -- ++(\letterspace, 0) node {$n_h$}
 -- ++(\letterspace, 0) node {$t$}; 
 \path[draw=white, line width = 3ex] (3.5, -5.5) node {$e$}
 -- ++(\letterspace, 0) node {$q$} 
 -- ++(\letterspace, 0) node {$u$} 
 -- ++(\letterspace, 0) node {$i_h$}
 -- ++(\letterspace, 0) node {$t$} 
 -- ++(\letterspace, 0) node {$t$}; 
 \path[draw=white, line width = 3ex] (3.9, -4.5) node {$e$}
 -- ++(\letterspace, 0) node {$q$} 
 -- ++(\letterspace, 0) node {$u$} 
 -- ++(\letterspace, 0) node {$t_h$} 
 -- ++(\letterspace, 0) node {$t$}; 
 \path[draw=white, line width = 3ex] (4.3, -3.5) node {$e$}
 -- ++(\letterspace, 0) node {$q$} 
 -- ++(\letterspace, 0) node {$u$} 
 -- ++(\letterspace, 0) node {$t_h$}; 
 \path[draw=white, line width = 3ex] (4.75, -2.5) node {$e_h$}
 -- ++(\letterspace, 0) node {$q$} 
 -- ++(\letterspace, 0) node {$u$}; 
 \path[draw=white, line width = 3ex] (5.5, -1.25) 
 -- ++(\letterspace, 0) node[draw=black, circle, inner sep=1.5, very thin] {$q_h$} 
 -- ++(\letterspace, 0) node {$u$}; 
 \path[draw=white, line width = 3ex] (8.5, -5.5) node {$u$} 
 -- ++(\letterspace, 0) node {$n$} 
 -- ++(\letterspace, 0) node {$i$}
 -- ++(\letterspace, 0) node {$q$} 
 -- ++(\letterspace, 0) node {$u$}
 -- ++(\letterspace, 0) node {$e_h$};
 \path[draw=white, line width = 3ex] (8, -4.5) node {$u$} 
 -- ++(\letterspace, 0) node {$n$} 
 -- ++(\letterspace, 0) node {$i_h$}
 -- ++(\letterspace, 0) node {$q$} 
 -- ++(\letterspace, 0) node {$u$};
 \path[draw=white, line width = 3ex] (7.5, -3.5) node {$u$} 
 -- ++(\letterspace, 0) node {$n_h$} 
 -- ++(\letterspace, 0) node {$q$}
 -- ++(\letterspace, 0) node {$u$};
 \path[draw=white, line width = 3ex] (7, -2.5) node {$u$} 
 -- ++(\letterspace, 0) node {$q_h$}
 -- ++(\letterspace, 0) node {$u$};
 \path[draw=white, line width = 3ex] (6.85, -1.25) 
 -- ++(\letterspace, 0) node[draw=black, circle, inner sep=1.5, very thin] {$u_h$} 
 -- ++(\letterspace, 0) node {$u$}
 -- ++(\letterspace, 0);
 \end{tikzpicture}
 \caption{\label{fig-ex-Vorder-gen}
 $unique \succ equitant$}
 \end{figure}
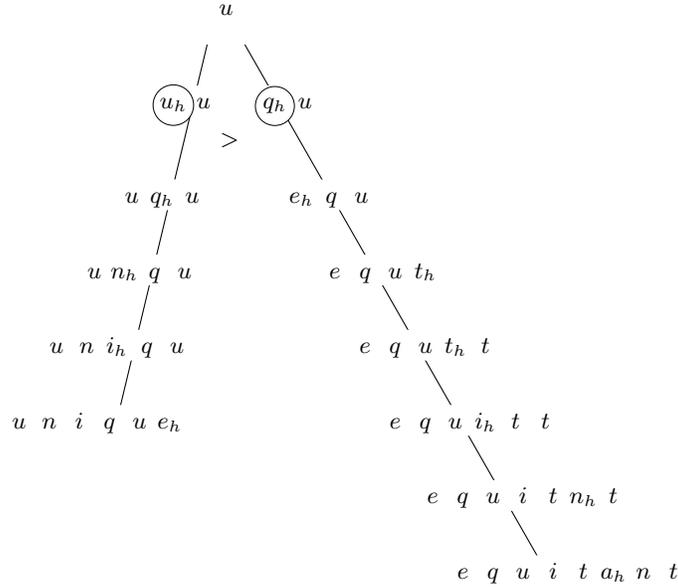
 \end{examp}

We now describe a canonical form of a string known as $V$-form which partitions a string according to its largest letter.

\begin{defn}[\itbf{$V$-form}~\cite{DaD96,DD03,DDS11,DDS13}]\label{Vform}
The \itbf{$V$-form} of any given string \s{x} is 
$$V_k(\s{x}) = \s{x} = \s{x_0}g\s{x_1}g\cdots\s{x_{k-1}}g\s{x_k},$$
where $g$ is the largest letter in \s{x} --- thus we suppose that $g$ occurs exactly $k$ times.
Note that any \s{x_i} may be the empty string $\s{\varepsilon}$.  We write $\mathcal{L}_{\s{x}}=g$,
$\mathcal{C}_{\s{x}}=k$.
\end{defn}

\begin{lemm}\label{lem-a}
{\cite{DaD96,DD03,DDS11,DDS13}} Suppose we are given distinct strings
$\s{x}$ and $\s{y}$ with corresponding $V$-forms
\begin{eqnarray}
\s{x} & = & \s{x}_0 \mathcal \EL_{\s{x}} \s{x}_1 \mathcal \EL_{\s{x}} \s{x}_2 \cdots
\s{x}_{j-1} \mathcal \EL_{\s{x}} \s{x}_{j}, \nonumber \\
\s{y} & = & \s{y}_0 \mathcal \EL_{\s{y}} \s{y}_1 \mathcal \EL_{\s{y}} \s{y}_2 \cdots
\s{y}_{k-1} \mathcal \EL_{\s{y}} \s{y}_{k}, \nonumber
\end{eqnarray}
where $j = \mathcal{C}_{\s{x}},\ k = \mathcal{C}_{\s{y}}$.
Let $h \in 0 .. \max(j,k)$ be the least integer such that $\s{x}_h
\neq \s{y}_h$. Then  $\s{x} \prec \s{y}$ if and only if one of the
following conditions holds:
\begin{description}
\item [(C1)] $\mathcal \EL_{\s{x}} < \mathcal \EL_{\s{y}}$\label{lem-a-c1}
\item [(C2)] $\mathcal \EL_{\s{x}} = \mathcal \EL_{\s{y}}$ and
$\mathcal{C}_{\s{x}} < \mathcal{C}_{\s{y}}$ \label{lem-a-c2}
\item [(C3)] $\mathcal \EL_{\s{x}} = \mathcal \EL_{\s{y}}$,
$\mathcal{C}_{\s{x}} = \mathcal{C}_{\s{y}}$
and $\s{x}_h \prec \s{y}_h$.\label{lem-a-c3} 
\end{description} 
\end{lemm}

\begin{lemm}\label{lem-subsequce}
{\cite{DDS11,DDS13}} For given strings $\s{x}$ and \s{y},
if \s{y} is a proper subsequence of \s{x},
then $\s{y}\prec \s{x}$. 
\end{lemm}

For instance, given $\s{x}=7547223$,
Lemma~\ref{lem-subsequce} states that $772 \prec 7547223$. 
Furthermore, another consequence of Lemma~\ref{lem-subsequce}  
is that shorter suffixes are suffixes of longer suffixes; that is, they occur from the shortest to the longest in increasing order. So for \s{x},
the $V$-order of the suffixes is 
just $3 \prec 23 \prec 223 \prec 7223 \prec 47223 \prec 547223 \prec 7547223$.
Thus in $V$-order suffix sorting is trivial,
in contrast to lexorder, where over the years numerous non-trivial (though linear) algorithms have been proposed \cite[ch.\ 5.2-5.3]{S03}.

Furthermore, string comparison is fast in $V$-order: 
\begin{lemm}
\label{claim:V-comp}
\cite{DDS11,DDS13,ADRS14,ADRS14FI,ADKRS16,ADM18} $V$-order comparison of given strings \s{x} and \s{y} requires linear time and constant space.
\end{lemm}

We now introduce the $V$-order equivalent of the lexorder Lyndon word:
\begin{defn}[{\bf $V$-word}~\cite{DD03}]\label{Vword}
A string \s{x} over an ordered alphabet $\Sigma$ is a \itbf{$V$-word} if it is the unique minimum in $V$-order $\prec$
over the conjugacy class of
\s{x}.
\end{defn}
Thus, like a Lyndon word, a $V$-word is necessarily primitive. 

\begin{examp}
\label{ex-1}
[$\prec$]
We can apply Definition \ref{def-V_order}, equivalently the methodology of Lemma \ref{lem-a}, to conclude that 
$$6263 \prec 6362 \prec 2636 \prec 3626,$$
so that $6263$ is a $V$-word,
while on the other hand $2636$ is a Lyndon word.
Similarly,
$62626263$ and $929493$ are $V$-words,
while conjugates $26262636$ and $294939$ are Lyndon words.
\end{examp}


We now define another important ordering:

\begin{defn} [{\bf $V$-word order}]
\label{Vworder}
Suppose \s{u} and \s{v} are $V$-words on an ordered alphabet $\Sigma$.
If \s{uv} is also a $V$-word, then we write $\s{u} <_{\mathcal{V}} \s{v}$;
if not, then 
$\s{u} \ge_{\mathcal{V}} \s{v}$.
\end{defn}
Thus, corresponding to the Lyndon factorization into Lyndon words
using $\ge$ (Theorem~\ref{lLyndon-thrm}),
we arrive at a {\it $V$-order factorization} expressed
in terms of $V$-word order $\ge_{\mathcal{V}}$:

\begin{lemm}
\label{claim:V-fact}
\cite{DDS11,DDS13,DBLP:journals/tcs/DaykinMS21}
[{\bf $V$-order Factorization}]
Using only linear time and space (see Algorithm VF in \cite{DDS13}\footnote{
VF is an on-line algorithm as it outputs the $V$-word factors in order from left to right without any backtracking. However, this fact was not explicitly stated in the reference.}),
a string \s{x} can be factored uniquely, using $V$-word order, into $V$-words
$\s{x} = \s{x_1x_2}\cdots\s{x_m}$,
where $\s{x_1}\ge_{\mathcal{V}}\s{x_2}\ge_{\mathcal{V}}\cdots \ge_{\mathcal{V}}\s{x_m}$.
\end{lemm}


\begin{examp}
\label{ex-2}
For $\s{x} = 33132421$, the Lyndon decomposition (computed using lexorder) is $3 \ge 3 \ge 13242 \ge 1$,
while the $V$-order factorization identifies nonextendible $V$-words 33132 and 421 with
$33132 \ge_{\mathcal{V}} 421$.
(Note however that $33132 \prec 421$!
See \cite{DDS09} for more background on this phenomenon.)
Similarly, from Example~\ref{ex-1},
the string
$$\s{x} = \s{uvw} = (6263)(62626263)(929493)$$
has the unique $V$-order factorization $\s{u} \ge_{\mathcal{V}} \s{v} \ge_{\mathcal{V}} \s{w}$,
even though $\s{u} \prec \s{v} \prec \s{w}$.
\end{examp}

It will also be useful to order strings \s{x}, \s{y}
based on a lexicographic approach to their 
factorizations into identified substrings; this will be applied in Section \ref{sec_Lyn-subset} to handle string factorization based not on letters but substrings.
We call this ordering, denoted 
$\prec_{LEX(F)}$, {\it lex-extension} order, expressed here with respect to substring
ordering using $\prec$ --- but note that other substring ordering methodologies could instead be applied. 

\begin{defn}[{\bf Lex-extension order (Lex-Ext order)} \cite{DD03,DDS13,DS14}]
\label{def-lex-ext}
Suppose that, according to some factorization $F$, two strings $\s{x}, \s{y} \in \Sigma^+$ are expressed in terms of nonempty factors: $$\s{x} = \s{x_1x_2}\cdots\s{x_m},\ \s{y} = \s{y_1y_2}\cdots\s{y_n}.$$ Then $\s{x} \prec_{LEX(F)} \s{y}$ if and only if one of the following holds:
\begin{itemize}
\item[(1)] $\s{x}$ is a proper prefix of $\s{y}$ (that is, $\s{x_i} = \s{y_i}$ for $1 \le i \le m < n$); or
\item[(2)]
for some least $i \in 1..\min(m,n)$,  $\s{x_j} = \s{y_j}$ for $j = 1,2,\ldots,i\- 1$, and $\s{x_i} \prec \s{y_i}$.
\end{itemize}
\end{defn}


We assume throughout that, when using lex-extension order, the factorization $F$ is given in $V$-form (Definition 
\ref{Vform}). That is, if the $V$-form of \s{x} is $$\s{x}  =  \s{x'_0} \mathcal \EL_{\s{x}} \s{x'_1} \mathcal \EL_{\s{x}} \s{x'_2} \cdots \s{x'_{k-1}} \mathcal \EL_{\s{x}} \s{x'_{k}},$$ 
then the corresponding 
lex-extension order factors are:
$$\s{x_1} = \s{x'_0},~ 
\s{x_2} = \mathcal \EL_{\s{x}}\s{x'_1},~ 
\cdots,~
\s{x_{k+1}} = \mathcal \EL_{\s{x}} \s{x'_{k}}.$$ 
Depending on the context, it could be that $\s{x'_0} = \s{\varepsilon}$, in which case $\s{x_1}  =  \mathcal \EL_{\s{x}}\s{x'_1}$, $\s{x_2}  =  \mathcal \EL_{\s{x}}\s{x}_2$, and so on.

Thus, in order to $V$-order two strings with identical $\EL$ values, 
we first compute the 
$V$-form factorization of each string, then treat each of the resulting factors as a single entry (a ``letter''), and so determine the ``lexicographic'' order of the given strings by comparing the factors from left to right using $\prec$ (Lemma \ref{lem-a} (C3) defines comparison of 
strings that are conjugates).


\section{UMFF and circ-UMFF theory}\label{sect-ummf}

Motivated by classical Lyndon words, investigations into combinatorial aspects of the factoring and concatenation of strings led to the concepts of UMFF and circ-UMFF \cite{DD03,DD08,DDS09},
whose properties we overview here and apply in Sections \ref{sec_Lyn-subset}, \ref{sect:sscircUMFF} and \ref{subsec_Galois}.

For given $\s{x} = \s{x}[1..n] \in \Sigma^+$, if $\s{x} = \s{w_1w_2\cdots w_k}$, $1 \le k \le n$,
then $\s{w_1w_2\cdots w_k}$
is said to be a \itbf{factorization} of \s{x};
moreover, if every factor \s{w_j}, $1 \le j \le k$, belongs to a specified
set $\mathcal{W}$,
then $\s{w_1w_2\cdots w_k}$
is said to be a \itbf{factorization
of \s{x} over $\mathcal{W}$}, denoted by $F_{\mathcal{W}}(\s{x})$.
A subset $\mathcal{W} \subseteq \Sigma^+$ is a
\itbf{factorization family} (FF) of $\Sigma$ if 
for every nonempty string
\s{x} on $\Sigma$ there exists a factorization $F_{\mathcal{W}}(\s{x})$. If for every $j = 1,2,...,k$, every factor $\s{w_j}$ is of maximum length, then the factorization $F_{\mathcal{W}}(\s{x})$ is unique and said to be \itbf{maximal}.

To show that not every factorization is necessarily maximal, consider the FF 
$\mathcal{W} = \{a,b,c,d,ab,cd,bcd\}$.
Then, for $\s{x} = abcd$, we get three possible factorizations, $(a)(b)(c)(d)$ and $(ab)(cd)$,  and also $(a)(bcd)$, depending on whether we process \s{x} in a forward or backward direction.  But therefore, in each case, not every factor can be of maximum length and so FF is not maximal.

{Observe that every FF must contain every element of $\Sigma$; moreover,
any subset of $\Sigma^+$ containing every element of $\Sigma$ is necessarily an FF.}

\begin{defn}
\label{defn-umff}
Let $\mathcal{W}$ be an FF on an alphabet $\Sigma$.
Then $\mathcal{W}$ is a {\bf unique maximal factorization family} (UMFF)
if and only if there exists a 
maximal factorization $F_{\mathcal{W}}(\s{x})$
for {\rm every} string $\s{x} \in \Sigma^+$.
\end{defn}

{The following characterization of UMFFs shows that there can be no overlapping factors in a unique maximal factorization of a string:}


\begin{lemm}
\label{lem-xyz}
(The $\mathbf{xyz}$ Lemma \cite{DD03})
An FF $\mathcal{W}$ is an {\bf UMFF} if and only if
whenever $\s{xy},\s{yz} \in \mathcal{W}$ for some nonempty \s{y},
then $\s{xyz} \in \mathcal{W}$.
\end{lemm}

Note that, although the Fibonacci words $b,a,ab,aba,abaab,abaababa\cdots$ are clearly an example of an FF, nevertheless they do not by Lemma~\ref{lem-xyz} constitute an UMFF: $\s{x} = abaab$, $\s{y} = aba$, $\s{z} = ab$ are all 
Fibonacci words,
as are $\s{xy}$ and $\s{yz}$, but $\s{xyz}$ is not. 

We next show that an FF that contains no overlapping factors --- such as the string $\s{y}$ in Lemma \ref{lem-xyz} --- necessarily forms an UMFF.

\begin{lemm}
\label{lem_y-empty}
Let $\mathcal{W}$ be an FF on $\Sigma$.  If for every distinct $\s{u,v} \in \mathcal{W}$ with $|\s{u}|, |\s{v}|>1$, \s{uv} is border-free,
then $\mathcal{W}$ is an UMFF.
\end{lemm}
\begin{proof}
By definition of FF, $\mathcal{W}$ must contain all the letters in $\Sigma$, which clearly do not overlap.  
Consider factoring some string $\s{x} = x_1 x_2 \ldots x_n$, $n>1$, maximally over $\mathcal{W}$, so that $\s{x} = \s{f_1}\s{f_2} \cdots \s{f_m}$, where 
no factor $\s{f_i}$, $1 \le i \le m$, can be extended either left or right. Specifically,
if $\s{f_i} = x_p \ldots x_q$, $1 \le p \le q \le n$, then $x_{p-j} \ldots x_{q+k} \notin \mathcal{W}$ for any positive $j < p,\ k < n+1-q$. 
The non-extendability of all factors ensures maximality and thus uniqueness of the factorization $\s{f_1}\s{f_2} \cdots \s{f_m}$, and so we conclude that $\mathcal{W}$ forms an UMFF.  \qed
\end{proof}
For example, with $\Sigma = \{0,1\}$, Lemma~\ref{lem_y-empty} tells us that $\mathcal{W}=\{0,1,010\}$ must be an UMFF.  On the other hand, the example FF
$\mathcal{W} = \{0,1,010,01010,0101010, \ldots\}$ from \cite{DD08}, with an infinity of bordered $\s{uv}$, is also an UMFF, showing that the converse of Lemma~\ref{lem_y-empty} certainly does not hold.
Indeed, observe from the following example that neither primitiveness nor the border-free property guarantees that a set of words forms an UMFF:

\begin{examp}
\label{ex_primitive}
Suppose that $\mathcal{W}$ is an UMFF over $\Sigma = \{a,b,c\}$ such that\\ $\{a,b,c, ab, abc, cab \} \subseteq \mathcal{W}$. Then consider applying the {\bf xyz} Lemma \ref{lem-xyz} twice to words in $\mathcal{W}$ as follows:
\begin{itemize}
\item[(i)] For $\s{x} = \s{z} = ab$ and $\s{y} = c$,  we find the bordered word $\s{xyz} = abcab \in \mathcal{W}$.
    \item[(ii)] For $\s{x} = abc$, $\s{z} = c$ and $\s{y} = ab$, we find the repetition $\s{xyz} =(abc)^2$ is also in $\mathcal{W}$.
\end{itemize}  
\end{examp}

Interestingly, known UMFFs in the literature, such as Lyndon words and $V$-words, are specified as being necessarily primitive and they also satisfy the border-free property.\\

An important class of UMFFs can now be specified:
\begin{defn}
\label{def-subcirc}
An UMFF $\mathcal{W}$ over $\Sigma^+$ is a {\bf circ-UMFF} if and only if
it contains exactly one rotation of every primitive string $\s{x} \in \Sigma^+$.
\end{defn}

Observe that the definition of UMFF does not require that $\Sigma$ be ordered.  Nor does the circ-UMFF $\mathcal {W}$, but it does require that the ordering of any two distinct strings in $\mathcal {W}$ depends only on their concatenation (that may or may not occur).
Thus a circ-UMFF $\mathcal {W}$ may be said to specifiy a ``\itbf{concatenation order}'':

\begin{defn}
\label{def-W-order}
(\cite{DDS09})
If a circ-UMFF $\mathcal {W}$ contains strings  $\s{u},\s{v}$ and \s{uv}, we write $\s{u} <_{\mathcal {W}} \s{v}$ (called {\bf $\mathcal{W}$-order)}.
\end{defn}

Observe that $V$-word order (Definition~\ref{Vworder}), also defined in terms of concatenation, is formally equivalent to
$\mathcal {W}$-order.

Structural properties of circ-UMFFs are summarized as follows:
\begin{thrm}
\label{thm-struct} 
(\cite{DD08}) Let $\mathcal{W}$ be a circ-UMFF.\\ 
(1) If $\s{u} \in \mathcal{W}$ then $\s{u}$ is
border-free.\\
(2) If $\s{u},\s{v} \in \mathcal {W}$ and $\s{u}
\neq \s{v}$ then $\s{uv}$ is primitive.\\ 
(3) If $\s{u},\s{v} \in
\mathcal {W}$ and $\s{u} \neq \s{v}$ then $\s{uv} \in \mathcal {W}$ or $\s{vu}
\in \mathcal {W}$ (but not both).\\
(4) If $\s{u},\s{v}, \s{uv} \in
\mathcal {W}$ 
then $\s{u} <_{\mathcal {W}} \s{v}$, 
and $<_{\mathcal {W}}$ is a
total order on $\mathcal {W}$.\\ 
(5) If $\s{w} \in \mathcal
{W}$ and $|\s{w}| \ge 2$ then there 
exist $\s{u},\s{v} \in
\mathcal {W}$ with $\s{w}=\s{uv}$. 
\end{thrm}
The first known circ-UMFF is believed to be the set of Lyndon words, 
whose specific $\mathcal {W}$-order is lexorder; that is, the usual ordering of the strings of $\Sigma^+$ is used to obtain Lyndon words.
{Formally, the Lyndon circ-UMFF applies the same lexicographic ordering as both $\Sigma^+$ and $\mathcal{W}$-order:}
\begin{thrm}
\label{thm-Duval} 
(\cite{Duval83}) Let $\mathcal{L}$ be the set of Lyndon words, and suppose $\s{u}, \s{v} \in \mathcal{L}$. Then $\s{uv} \in \mathcal{L}$ if and only if \s{u} precedes \s{v} in lexorder.
\end{thrm}

Note that, from Definition~\ref{defn-umff}, $V$-order factorization determines an UMFF, which, by Definitions~\ref{Vword} and \ref{def-subcirc}, is a circ-UMFF.


\section{$V$-words, Lyndon Words and circ-UMFFs}
\label{sec_Lyn-subset}

In this section we investigate further the relationship and differences between Lyndon and $V$-words and introduce generalized words over any total order.

We begin with an observation made 
in \cite{DDS09}, that follows immediately from Duval's fundamental Theorem \ref{thm-Duval} \cite{Duval83}:


\begin{obs}
\label{obs-subclasses}
Let $\Sigma^*_{lex}$ denote the lexicographic total ordering of $\Sigma^*$. Then the lexordered set $\mathcal{L}$ of Lyndon words is a 
suborder of $\Sigma^*_{lex}$.
\end{obs}
However, observe that there is no corresponding architecture for $V$-words.  In $V$-ordered $\Sigma^*$, for $\s{x} = 21, \s{y} = 31$, we have $\s{x} \prec \s{y}$ by Lemma \ref{lem-a} (C1), while in the class of $V$-words we have 
$\s{x} \ge_{\mathcal{V}} \s{y}$ by Definition \ref{Vworder} of $V$-word order. For further details on the distinction between $\prec$ and $\ge_{\mathcal{V}}$ see Lemma 3.16 in \cite{DDS13}.

Lyndon words and $V$-words are generally distinct \cite{DDS13}. For instance, the integer string $1236465123111$ factors into Lyndon words $(12316465)(123)(1)(1)(1)$ and into $V$-words $(1)(2)(3)(6465123111)$ --- no correspondence whatsoever. Nevertheless,
when substrings are restricted to a single letter, a rather remarkable result holds, which is a newly observed special case of Theorem 4.1 in \cite{DD03} and leads to the concept of $V$-Lyndons:

\begin{lemm}[$V$-Lyndons]
\label{lem-singletons}
Suppose \s{x} has a $V$-form $\s{x}  =  \mathcal \EL_{\s{x}} \s{x}_1 \mathcal \EL_{\s{x}} \s{x}_2 \cdots \s{x}_{j-1} \mathcal \EL_{\s{x}} \s{x}_{j}$, 
where $\s{x}_0 = \s{\varepsilon}$ and 
$|\s{x}_{l}| = 1$ for $1 \le l \le j$.
Let 
$ \s{x'} = \s{x}_1 \s{x}_2 \cdots \s{x}_{j-1} \s{x}_{j}$.  Then
\s{x} is a $V$-word if and only if \s{x'}  is a Lyndon word.
\end{lemm}
To see that the requirement 
$|\s{x}_{l}|=1$ is necessary,
consider $\s{x} = 321312$ with
$\EL_{\s{x}} = 3$, 
$|\s{x}_{1}| = |\s{x}_{2}|= 2$.
Certainly 
$\s{x'} = \s{x}_{1}\s{x}_{2} = 2112$ is 
\itbf{not} a Lyndon word, but since $\s{x} \prec 312321$,
$\s{x}$ is a $V$-word.  Thus Lemma \ref{lem-singletons} does not generalize to $V$-form substrings with 
$|\s{x}_{l}| > 1$.
Nonetheless, there does exist a kind of reciprocity between infinite classes of Lyndon words and $V$-words:
\begin{obs}
\label{lem-lyn2V}
 For any Lyndon word $\s{x}[1 .. n]$, $n \ge 2$, on ordered alphabet $\Sigma$:
\begin{itemize}
\item[(1)] If 
$\mathcal \EL_{\s{x}}$ is the largest letter in \s{x}, then $\mathcal (\EL')^k\s{x}$ is a $V$-word for $\mathcal \EL' > \mathcal \EL_{\s{x}}$ and every integer $k > 0$.
\item[(2)] If  
$\mathcal {\s{\ell}_{\s{x}}}$ is the smallest 
letter in \s{x},
then $\mathcal (\s{\ell}_{\s{x}})^k\s{x}$ is a Lyndon word for every integer $k > 0$.
\end{itemize}
\end{obs}


Building on Lemma \ref{lem-singletons} and Observation \ref{lem-lyn2V}(1), we can show that there are infinitely more $V$-words 
than there are Lyndon words: 
\begin{thrm}
\label{thm-infinite}
Suppose that $\s{\ell}[1 .. n]$ is a Lyndon word over an ordered alphabet $\Sigma$ and further that there exists $\mathcal \EL_{\s{\ell}}\in \Sigma$ such that $\mathcal \EL_{\s{\ell}} > \s{\ell}[i]$ for every $i \in 1 .. n$. 
Then we can construct infinitely many $V$-words from $\s{\ell}$ over $\Sigma$.
\end{thrm}

\begin{proof}
For the first $V$-word $\s{v}_1$, applying  
Lemma \ref{lem-singletons},
we rewrite $\s{\ell}$ as $\s{v}_1 [1 .. 2n]$ where for $i \in 1 .. 2n$, if $i$ is odd, $\s{v}_1[i] = \mathcal \EL_{\s{\ell}}$, while if $i$ is even, $\s{v}_1[i] = \s{\ell}[i/2]$; that is, $\s{v}_1 = \mathcal \EL_{\s{\ell}}\s{\ell}[1] \mathcal \EL_{\s{\ell}} \s{\ell}[2] .. \mathcal \EL_{\s{\ell}} \s{\ell}[n]$.

For $V$-words $\s{v}_h$, $h > 1$, rewrite $\s{\ell}$ as $\s{v}_h = \mathcal \EL_{\s{\ell}}\s{\ell}[1]^h \mathcal \EL_{\s{\ell}} \s{\ell}[2]^h .. \mathcal \EL_{\s{\ell}} \s{\ell}[n]^h$. Lemma \ref{lem-a} (C1) shows that if $a \prec b$ for letters $a,b$ (that is, $a<b$ in $\Sigma$), then $a^h \prec b^h$ and hence the Lyndon property (Lemma~\ref{lem-singletons}) of $\s{\ell}$ is preserved for $\s{v}_h$ using Definition \ref{def-lex-ext} for lex-extension order of strings.
\qed
\end{proof}

It might then be natural to suppose that $V$-words exhibit the same structural properties as Lyndon words and support equivalent string operations. For instance, a defining property of Lyndon words is that they are strictly less in lexorder than any of their proper suffixes; that is, for a Lyndon word $\s{\ell} = \s{p_{\ell}}\s{s_{\ell}}$, with $\s{p_{\ell}},\s{s_{\ell}} \neq \s{\epsilon}$, we have 
$$\s{\ell} < \s{s_{\ell}} < \s{s_{\ell}}\s{p_{\ell}}.$$
This central Lyndon property relates to two important operations on strings: ordering and concatenation.  For Lyndon words, these operations
are consistent with respect to lexorder: that is,  for every proper suffix $\s{s_{\ell}}$ of $\s{\ell}$, by virtue of the ordering $\s{\ell} < \s{s_{\ell}}$,
we can construct a Lyndon word  $\s{\ell}\s{s_{\ell}}^h$ by concatenation for every $h \ge 1$.  

In contrast, for $V$-order, these operations are not necessarily consistent.  First, by Lemma \ref{lem-subsequce}, a proper suffix \s{u} of a string \s{x}
is less than \s{x} in $V$-order; thus, 
for example, given the $V$-word
$\s{v} = 43214123$, even though substrings $23 \prec \s{v}$ and $4123 \prec \s{v}$, on the other hand, by definition of a $V$-word, $\s{v} \prec 41234321$. 
So for a $V$-word $\s{v} = \s{p_v}\s{s_v}$, with $\s{p_v},\s{s_v} \neq \s{\epsilon}$, 
we have $$\s{s_v} \prec \s{v} \prec \s{s_vp_v}.$$

Nevertheless, like a Lyndon word, a $V$-word can be concatenated with any of its proper suffixes (although they are less in $V$-order) to form a larger $V$-word (Lemma 3.21 in \cite{DDS13}). 
Hence we are interested in those combinatorial properties related to operations like concatenation and indexing in conjugacy classes which hold both for Lyndon words and $V$-words. Examples include border-freeness, existence of $\s{uv}$ and $\s{vu}$ in the conjugacy class where \s{u} and \s{v} are Lyndon words, and the FM-index Last First mapping property~\cite{DBLP:journals/tcs/DaykinMS21,FM00}. 

{More generally, it is intriguing to explore similarities and differences between instances of circ-UMFFs, as discussed below.} 
We begin by introducing a general form of order, $\mathcal{T}$:
\begin{defn}
\label{def_Torder}
($\mathcal{T}$-order)
Let $\mathcal{T}$ be any total ordering of $\Sigma^*$ with order relation $\ll$ so that given distinct strings \s{x}, \s{y} they can be ordered deterministically with the {\bf relation $\ll$}: either $\s{x} \ll \s{y}$ or $\s{y} \ll \s{x}$. 
\end{defn}
So for Lyndon words ($V$-words) the ordering $\mathcal{T}$ is lexorder ($V$-order) and the corresponding order relation $\ll$ is $<$ ($\prec$).  Using the general order $\mathcal{T}$, we can extend Definitions~\ref{def-lyndon}/\ref{Vword} from Lyndon words/$V$-words to $\mathcal{T}_{\ll}$-words:
\begin{defn}
\label{def_Tword}
($\mathcal{T}_{\ll}$-word)
A string \s{x} over an ordered alphabet $\Sigma$ is said to be a {\bf \itbf{$\mathcal{T}_{\ll}$-word}} if it is the unique minimum in $\mathcal{T}$-order $\ll$
in the conjugacy class of \s{x}.
\end{defn}
Similarly, Definition \ref{def-lex-ext} (lex-extension order) can be generalized by replacing the order $\prec$ by $\mathcal{T}$-order $\ll$:


\begin{defn}[$\mathcal{T}_{lex}$-order] 
\label{Torder}
For a factorization $F$, let $\s{x}, \s{y} \in \Sigma^+$ be two strings expressed in terms of nonempty factors: $$\s{x} = \s{x_1x_2}\cdots\s{x_m},\ \s{y} = \s{y_1y_2}\cdots\s{y_n}.$$ Then $\s{x} \ll_{LEX(F)} \s{y}$ if and only if:
\begin{itemize}
\item[(1)] $\s{x}$ is a proper prefix of $\s{y}$ or
\item[(2)]
for some least $i \in 1..\min(m,n)$,  $\s{x_j} = \s{y_j}$ for $j = 1,2,\ldots,i\- 1$, and $\s{x_i} \ll \s{y_i}$.
\end{itemize}
\end{defn}


Applications of circ-UMFFs in the literature arise in linear-time variants of the Burrows-Wheeler transform: the $V$-order based transform $V-BWT$ \cite{DS14}; the binary Rouen transform $B-BWT$ derived from binary block order which generated twin transforms \cite{DBLP:journals/tcs/DaykinGGLLLP16}; the degenerate transform $D-BWT$ for indeterminate strings implemented with lex-extension order \cite{DaykinW17} which supports backward search \cite{DaykinGGLLLMPW19}. These instances stimulate the quest for new circ-UMFFs and we pave the way for this by next introducing a generalization of circ-UMFFs.

\section{Substring circ-UMFF: Generalization of circ-UMFF}\label{sect:sscircUMFF}

Definitions~\ref{def_Torder} and \ref{def_Tword} encourage considering conjugacy classes for substrings 
rather than individual letters;
that is, each 
conjugate is defined by a rotation with prefix $\mathcal \EL_{\s{x}}\s{x_i}$, 
as follows:



\begin{defn}
\label{defn-substring_conjugate}
Suppose that a string $\s{x} = \s{x}[1..n]$
over an ordered alphabet $\Sigma$, with maximal letter $\mathcal \EL_{\s{x}}$,
is expressed
in $V$-form as
$\mathcal \EL_{\s{x}} \s{x}_1 \mathcal \EL_{\s{x}} \s{x}_2 \cdots \s{x}_{j-1} \mathcal \EL_{\s{x}} \s{x}_{j}$,
with $\s{x}_0 = \s{\varepsilon}$. 
Then for every 
$1 \le t \le j$,
$$\s{y} = \mathcal{R}_t(\s{x}) = \EL_{\s{x}} \s{x}_{t+1} \mathcal \EL_{\s{x}} \s{x}_{t+2} \cdots \s{x}_{j-1} \mathcal \EL_{\s{x}} \s{x}_{j}(\EL_{\s{x}} \s{x}_1 \cdots \mathcal \EL_{\s{x}}\s{x}_{t})$$ is the $t^{th}$ {\bf substring conjugate} (or {\bf substring rotation}) of $\s{x}$ 
(so $\mathcal{R}_j(\s{x}) = \s{x}$).
\end{defn}




Since Lemma \ref{lem-xyz} holds for unrestricted strings \s{xy} and \s{yz} (given \s{y} is nonempty) then it certainly holds for specified types of substrings, so in the context of substring conjugates we get:

\begin{obs}
\label{obs-substring_xyz}
Lemma \ref{lem-xyz} holds for substrings expressed in V-form.    
\end{obs}


Thus a natural generalization of circ-UMFF is the {\it substring circ-UMFF}, where a conjugate is selected from the conjugacy class of substrings of a string rather than the usual rotation of letters. 

\begin{defn}
\label{def-subcircUMFF}
An UMFF $\mathcal{W}$ over $\Sigma^+$ is a \itbf{substring circ-UMFF} if and only if
it contains exactly one substring rotation of every primitive string $\s{x} \in \Sigma^+$ expressed in $V$-form.
\end{defn}


\begin{defn}\label{lexext-word}
A string $\s{x}  =  
\mathcal \EL_{\s{x}} \s{x}_1 \mathcal \EL_{\s{x}} \s{x}_2 \cdots \s{x}_{j-1} \mathcal \EL_{\s{x}} \s{x}_{j}$ in $V$-form 
over an ordered alphabet $\Sigma$, with maximal letter $\mathcal \EL_{\s{x}}$, is said to be a 
{\bf \itbf{$\mathcal{T}_{lex}$-word}}
if it is the unique minimum in $\mathcal{T}_{lex}$-order 
in its conjugacy class. 
\end{defn}


To clarify, consider the primitive integer string 431412 where, with reference to $V$-form, $\mathcal \EL = 4$, and letting $\ll$ denote co-lexorder (lexorder of reversed strings), then the conjugate 412431 is least in co-lexorder for the letter-based conjugates, while the substring conjugate 431412 is least in Lex-Ext co-lexorder in the comparison of 431412 and 412431. That is, 412431 is a \itbf{$\mathcal{T}_{\ll}$-word}, while 431412 is a \itbf{$\mathcal{T}_{lex}$-word}.    


We have then the following important result:
\begin{thrm}
\label{thm-newcircUMFF}
Suppose $\ll$ is a $\mathcal{T}$-order over $\Sigma^*$.
(i) The class of border-free
$\mathcal{T}_{\ll}$-words forms a circ-UMFF $\boldsymbol{\mathcal{T}}$
over the conjugacy class of letters.
(ii) The class of border-free
$\mathcal{T}_{lex}$-words
forms a substring circ-UMFF 
over the conjugacy class of substrings.
\end{thrm}
\begin{proof}
By Definition~\ref{def_Torder}, $\ll$ is a total order.
Here we make use of two fundamental results: the \s{xyz} lemma (Lemma~\ref{lem-xyz} from \cite{DD03}) and the circ-UMFF theorem (Theorem~\ref{thm-struct} from \cite{DD08}).

{\itbf {Part (i)}}. Let $\boldsymbol{\mathcal{T}}$ denote the set of border-free $\mathcal{T}_{\ll}$-words (Definition \ref{def_Tword}) over $\Sigma$.
First, by the definition of $\boldsymbol{\mathcal{T}}$, every letter in $\Sigma$ is in $\boldsymbol{\mathcal{T}}$, so we may confine our consideration to strings of non-unit length.

Suppose then that $\s{xy}$ and $\s{yz}$, with \s{x,y,z} nonempty, are both border-free $\mathcal{T}_{\ll}$-words in $\boldsymbol{\mathcal{T}}$, therefore primitive.
Consider the string $\s{xyz}$ and suppose that it is a repetition $\s{u}^k$, $k>1$. 
But if $\s{u}$ is a proper prefix of \s{xy} then \s{xy} is bordered, whereas if \s{xy} is a prefix of $\s{u}$ then \s{yz} is bordered.  Thus we conclude that $\s{xyz}$ is also primitive, and so 
must have at least one border-free conjugate --- 
for instance, the conjugate
that is a Lyndon word is border-free and therefore might be 
in $\boldsymbol{\mathcal{T}}$. So we next
consider which border-free conjugate $\s{c}_{\mathcal{T}}$ of \s{xyz} is in $\boldsymbol{\mathcal{T}}$.

So suppose that $\s{xyz}$ is not itself minimal in $\mathcal{T}$-order in its conjugacy class.
Let $\s{x} = x_1 x_2 \ldots x_r$, $\s{y} = y_1 y_2 \ldots y_s$ and $\s{z} = z_1 z_2 \ldots z_t$, $r,s,t \ge 1$.  First assume that a conjugate $\s{c}=x_{c+1} \ldots x_r \s{y} \s{z} x_1 \ldots x_c$, 
$1 \le c \le r$,
is minimal, thus border-free and in $\boldsymbol{\mathcal{T}}$. But then applying Lemma \ref{lem-xyz} to $\s{c}$ and $\s{xy}$ implies that the bordered word $x_{c+1} \ldots x_r \s{y} \s{zxy}$ is in $\boldsymbol{\mathcal{T}}$, an impossibility. So assume that a conjugate $\s{c'}=y_{d+1} \ldots y_s \s{z} \s{x} y_1 \ldots y_d$, 
$1 \le d \le s$ is minimal and belongs to $\boldsymbol{\mathcal{T}}$. Again applying Lemma \ref{lem-xyz} to $\s{yz}$ and $\s{c'}$ 
implies that the bordered word 
$\s{yzx}y_1 \ldots y_d$ is in $\boldsymbol{\mathcal{T}}$, again impossible. Finally, for 
$\s{c''}=z_{e+1} \ldots z_t \s{x} \s{y} z_1 \ldots z_e$, $e+1 \ge 1$, 
a similar argument for $\s{c''}$ and $\s{yz}$ implies that the bordered word $z_{e+1} \ldots z_t \s{xyz}$ is in $\boldsymbol{\mathcal{T}}$.   
Hence the primitive conjugate \s{xyz} must itself be border-free and so must be the one, $\s{c}_{\mathcal{T}}$, in $\boldsymbol{\mathcal{T}}$, 
that is least in $\mathcal{T}$-order $\ll$ in its conjugacy class. Applying the sufficiency in Lemma \ref{lem-xyz}, since $\s{xy}$, $\s{yz}$ and $\s{xyz}$ with nonempty \s{y} all belong to $\boldsymbol{\mathcal{T}}$, we can conclude that $\boldsymbol{\mathcal{T}}$ is 
an UMFF. 

Thus, from each conjugacy class of a primitive string, we have shown how to select a border-free word for $\boldsymbol{\mathcal{T}}$, therefore satisfying Definition \ref{def-subcirc}.
Since moreover circ-UMFFs are necessarily border-free 
(Theorem \ref{thm-struct} Part (1)), we can conclude that $\boldsymbol{\mathcal{T}}$ is a circ-UMFF.\\

\itbf{Part (ii)}. The proof here is similar to that of Part (i), substituting Definition \ref{lexext-word} of $\mathcal{T}_{lex}$-words for
Definition \ref{def_Tword} of $\mathcal{T}_{\ll}$-words, and applying Observation \ref{obs-substring_xyz} for substrings.
Here we let $\boldsymbol{\mathcal{T^s}}$ denote the set of border-free $\mathcal{T}_{lex}$-words over $\Sigma^*$.
Then by the definition of $\boldsymbol{\mathcal{T^s}}$, the substrings of unit length (``letters") are in $\boldsymbol{\mathcal{T^s}}$, where
these strings of length one have the form $\mathcal{L}\s{w_i}$ where $\s{w_i} \in \Sigma^*$ (so if $\s{w_i}$ is the empty string we get the 
substring $\mathcal{L}$). Then a non-unit length string \s{w} of length $\ell$ will have $\ell$ occurrences of $\mathcal{L}$ in the form $\mathcal \EL_{\s{w}} \s{w_1} \mathcal \EL_{\s{w}} \s{w_2} \cdots \s{w_{\ell-1}} \mathcal \EL_{\s{w}} \s{w_{\ell}}$. 
Suppose then that $\s{xy}$ and $\s{yz}$, with \s{x,y,z} nonempty, are both border-free $\mathcal{T}_{lex}$-words in $\boldsymbol{\mathcal{T^s}}$, therefore primitive, and as above we deduce that \s{xyz} is also primitive.
In the analysis of $\boldsymbol{\mathcal{T^s}}$, for the existence of at least one border-free substring conjugate, 
we observe that the $V$-order 
circ-UMFF can be considered in the form of Definition \ref{def-subcircUMFF}. Then there exists a border-free 
$V$-word in the substring conjugacy class. As above, we suppose that \s{xyz} is not minimal in $\mathcal{T}_{lex}$-order and proceed to 
consider which border-free conjugate $\s{c}_{\mathcal{T^s}}$ of primitive \s{xyz} is in $\boldsymbol{\mathcal{T^s}}$. Let $\s{x} = \mathcal \EL_{\s{x}} \s{x}_1 \mathcal \EL_{\s{x}} \s{x}_2 \cdots \s{x}_{r-1} \mathcal \EL_{\s{x}} \s{x}_{r}$, $\s{y} = \mathcal \EL_{\s{y}} \s{y}_1 \mathcal \EL_{\s{y}} \s{y}_2 \cdots \s{y}_{s-1} \mathcal \EL_{\s{y}} \s{y}_{s}$, $\s{z} = \mathcal \EL_{\s{z}} \s{z}_1 \mathcal \EL_{\s{z}} \s{z}_2 \cdots \s{z}_{t-1} \mathcal \EL_{\s{z}} \s{z}_{t}$, $r,s,t \ge 1$. First assume that a conjugate 
$\s{c} = \mathcal \EL_{\s{x}} \s{x_{c+1}} \ldots \mathcal \EL_{\s{x}} \s{x_r} \s{y} \s{z} \mathcal \EL_{\s{x}}\s{x_1} \ldots \mathcal \EL_{\s{x}} \s{x_c}$, 
$1 \le c \le r$,
is minimal, thus border-free and in $\boldsymbol{\mathcal{T^s}}$. But then applying
Observation \ref{obs-substring_xyz}
to 
$\s{c}$ and $\s{xy}$ implies that the bordered word $\mathcal \EL_{\s{x}} \s{x_{c+1}} \ldots \mathcal \EL_{\s{x}}\s{x_r} \s{y} \s{zxy}$ is in $\boldsymbol{\mathcal{T^s}}$, an impossibility. 
The rest of the argument follows as in Part (i), first showing that \s{xyz} is the conjugate $\s{c}_{\mathcal{T^s}}$ in $\boldsymbol{\mathcal{T^s}}$, and finally establishing that $\boldsymbol{\mathcal{T^s}}$ is a substring circ-UMFF.
\qed
\end{proof}

Observe that Theorem \ref{thm-newcircUMFF} shows that if a circ-UMFF or a substring circ-UMFF is defined using a total order (which is not necessary), then every element of the (substring) circ-UMFF is obtained using the same total order and no other ordering technique. Observe further that the proof does not depend on any particular method of totally ordering $\Sigma^*$; however, the method must be total for border-free (hence primitive) strings in $\Sigma^*$.

We illustrate concepts from Theorem \ref{thm-newcircUMFF} with the following:

\begin{examp}
Consider the border-free integer string $\s{x} = 3177412$. Then the (unordered) conjugacy class of \s{x} is given by
\label{ex-conjugates}
    \begin{center}
     \[\begin{matrix} 
	3 & 1 & 7 & 7 & 4 & 1 & 2 \\
	2 & 3 & 1 & 7 & 7 & 4 & 1 \\
	1 & 2 & 3 & 1 & 7 & 7 & 4 \\
	4 & 1 & 2 & 3 & 1 & 7 & 7 \\
	7 & 4 & 1 & 2 & 3 & 1 & 7 \\
	7 & 7 & 4 & 1 & 2 & 3 & 1 \\
	1 & 7 & 7 & 4 & 1 & 2 & 3 
    \end{matrix}\]
	\end{center}
	The third conjugate, $1231774$, is the Lyndon word as it is least in lexorder. The sixth conjugate, $7741231$, is the $V$-word as it is least in $V$-order; it is also a co-lexorder word as it is least in co-lexorder; furthermore, it is least in relex order (reverse lexorder). The fifth conjugate, $7412317$, is the second largest in lexorder but is a bordered word. And the seventh conjugate, $1774123$,  is least in alternating lexorder (indexing strings from 1, odd indexed letters are compared with $<$ and even with $>$).
\end{examp}



We illustrate Definition \ref{def-lex-ext} for Lex-Ext co-lexorder where we intertwine lexorder 
with ordering substrings in co-lexorder:
\begin{examp}
\label{ex-colyn}
This example establishes the Lex-Ext co-lexorder substring circ-UMFF. Given a string $\s{x} \in \Sigma^*$  in $V$-form, $\s{x}  =  \s{x}_0 \mathcal \EL_{\s{x}} \s{x}_1 \mathcal \EL_{\s{x}} \s{x}_2 \cdots \s{x}_{j-1} \mathcal \EL_{\s{x}} \s{x}_{j}$, 
with $\s{x}_0 = \s{\varepsilon}$, 
the substring conjugates of \s{x} are compared using lex-extension where the $\mathcal \EL \s{x}_i$ substrings will be compared in co-lexorder. 
Since both lexorder and by isomorphism co-lexorder are total orders of $\Sigma^*$, any pair of distinct strings $\s{x}, \s{y} \in \Sigma^*$ can be compared deterministically in Lex-Ext co-lexorder. So given a border-free, hence primitive, string $\s{x}$, we can uniquely choose a conjugate from the substring conjugacy class which is minimal in Lex-Ext co-lexorder ({\it cf.} Definition \ref{def-subcirc}).  Theorem 
\ref{thm-newcircUMFF} Part (ii) then applies where in this case the class of border-free  
$\mathcal{T}_{lex}$-words is the set of Lex-Ext co-lexorder words. 
For example, the integer string 
9211912197194395119119111912 factors uniquely as (921191219719439511)(9119111912),
while into Lex-Ext lexorder words as (9211)(921951)(9119111912).
\end{examp}
Clearly we can replace co-lexorder in this context with any method for totally ordering $\Sigma^*$, for instance relex order or 
alternating lexicographic order. Note that, as shown for $V$-order \cite{DDS13}, co-lexorder string 
concatenation and ordering are not necessarily the same --- this phenomenon of circ-UMFFs is explored in \cite{DDS09}. For example, 321 is less than 54 in co-lexorder, while 54 is less than 321 in circ-UMFF order, because the concatenation 54321 is a co-lexorder word.

Naturally, we can modify the results of this section by defining other canonical forms of a string.  For instance, rather than partitioning a string according to the maximal letter $\mathcal \EL$ as in $V$-form, the partition could depend on the minimal letter, or indeed any well-defined substring pattern --- such as a short palindromic motif in DNA sequences.

Implementing the FM-Index in $V$-order was considered in \cite{DBLP:journals/tcs/DaykinMS21}, leading to $V$-order substring pattern matching using backward search --- whereby computing only on the $k$ conjugates starting with the greatest letter, essentially a substring circ-UMFF, reduced the BWT matrix to $O(nk)$ space. Note, however, that to fully implement BWT-type pattern matching in $V$-order so as to handle all letters --- rather than just $V$-letters which have only one maximal letter which occurs at the start of the substring --- 
remains an open problem \cite{DBLP:journals/tcs/DaykinMS21}.
 Hence, the substring circ-UMFF concept promises future optimization opportunities --- in particular, related to indexing and pattern matching applications.

\section{Galois words}
\label{subsec_Galois}

This final section applies many of the concepts 
developed in Sections \ref{sect-ummf}--\ref{sect:sscircUMFF} to show that although --- in contrast to circ-UMFF words and in particular to classic Lyndon words --- a given string does not necessarily factor uniquely and maximally into Galois words, nevertheless Galois words may still belong to a set of words which does form a circ-UMFF. 
Additionally, the continuing necessity of the border-free requirement, as in Theorem \ref{thm-newcircUMFF}, is demonstrated. We first 
describe Galois words as
introduced in \cite{R06}, based on a variant of lexorder, namely alternating lexicographic order, denoted $\prec_{alt}$ --- in which,
informally, positions within two given strings are compared in alternating $<$ and $>$ order. More precisely:

\begin{defn} 
\label{def_alt_lexorder}
[\itbf{Alternating lexorder $\prec_{alt}$ (modified from \cite{DBLP:journals/tcs/GiancarloMRRS20})}]
Given distinct strings $\s{x} = x_1 x_2 \ldots x_{s}$ and $\s{y} = y_1 \ldots y_{t}$, 
$1 \le s \le t$:\\ 
(1) (\s{x} not a proper prefix of \s{y}) If $i$ is the smallest index such that $x_i \neq y_i$, 
then $\s{x} \prec_{alt} \s{y}$ iff 
(a) $i$ is odd and $x_i < y_i$ or
(b) $i$ is even and $y_i < x_i$. 
Otherwise, $\s{y} \prec_{alt} \s{x}$.\\
(2) (\s{x} a proper prefix of \s{y})
$\s{x} \prec_{alt} \s{y}$ iff $|\s{x}|$ is even.
Otherwise, $\s{y} \prec_{alt} \s{x}$.
\end{defn}

An immediate consequence of Definition~\ref{def_alt_lexorder}(2) is the following:

\begin{lemm}
    For any string  $\s{x}$, if $|\s{x}|$ is even then 
    (1) $\forall \s{y} \in \Sigma^+$, $\s{x} \prec_{alt} \s{xy}$; in particular,
    (2) $\s{x}^k \prec_{alt} \s{x}^{k+r}$, for $k \geq 1, r \geq 1$.
\end{lemm}
We first establish that alternating lexorder $\prec_{alt}$ forms a total order over $\Sigma^*$, thus enabling the selection of a unique conjugate (such as the least) for Definition \ref{def-subcirc} and as also required for Theorem 
\ref{thm-newcircUMFF}. 


\begin{thrm}
\label{thm-altorder}
Alternating lexorder $\prec_{alt}$ is a total order over $\Sigma^*$.
\end{thrm}
\begin{proof} 
Consider distinct $\s{x}, \s{y}, \s{z} \in \Sigma^+$ with $\s{x} = x_1x_2 \ldots x_e$, $\s{y} = y_1y_2 \ldots y_f$, $\s{z} = z_1z_2 \ldots z_g$. 
Assume that $\s{x} \prec_{alt} \s{y} \prec_{alt} \s{z}$, but that $\s{z} \prec_{alt} \s{x}$.\\
{\bf Case 1.} $\s{z}$ is a proper prefix of $\s{x}$.
Since 
$\s{z} \prec_{alt} \s{x}$, by Definition \ref{def_alt_lexorder}(2) $|\s{z}|$ is even.\\
{\bf Subcase 1a.} $\s{x}$ is a proper prefix of $\s{y}$. 
Thus $\s{z}$ must be a proper prefix of $\s{y}$, and since $|\s{z}|$ is even,
again by Definition \ref{def_alt_lexorder}(2)
$\s{z} \prec_{alt} \s{y}$, a contradiction.\\
{\bf Subcase 1b.} $\s{x}$ is not a proper prefix of $\s{y}$. Then there is a smallest index $i$ such that $x_i \neq y_i$. If $i > |\s{z}|$, then \s{z} is a proper prefix of \s{y} and the contradiction of Subcase 1a applies to \s{z} and \s{y}. So suppose $i \le |\s{z}|$ and consider a corresponding relation $R$, 
$<$ or $>$ for odd or even index $i$, respectively --- then since $z_i = x_i$, so that $x_i~R~ y_i$ implies $z_i~R~ y_i$, it follows that $\s{z} \prec_{alt} \s{y}$, a contradiction.\\
{\bf Case 2.} $\s{z}$ is not a proper prefix of $\s{x}$. \\
{\bf Subcase 2a.} If $\s{x}$ is a proper prefix of $\s{y}$, then $\s{z} \prec_{alt} \s{y}$, a contradiction.\\ 
{\bf Subcase 2b.} If $\s{x}$ is not a proper prefix of $\s{y}$, then there is a smallest index $j$ such that $x_j \neq y_j$ and $x_j ~R~ y_j$ (where $R$ is as above based on odd/even index).  Let $i$ be the smallest index such that $z_i \neq x_i$.  If $j<i$, then $z_j = x_j$ and $x_j ~R~ y_j$,
so that $\s{z} \prec_{alt} \s{y}$, a contradiction. If $j>i$, then 
$z_i ~R~ x_i = y_i$
and $\s{z} \prec_{alt} \s{y}$, also a contradiction. If $j=i$, then 
$z_j ~R~ x_j ~R~ y_j$ and $\s{z} \prec_{alt} \s{y}$, again a contradiction. \\ \\
We conclude that $\s{x} \prec_{alt} \s{y} 
\prec_{alt} \s{z}$ 
requires $\s{x} \prec_{alt} \s{z}$; 
that is, the transitivity establishes that $\prec_{alt}$ is a total order over $\Sigma^+$.
As usual, $\s{\varepsilon}$ is the unique least string, and so $\prec_{alt}$ forms a total order over $\Sigma^*$.
\qed
\end{proof}





Since $\prec_{alt}$ is a total order, it is a candidate for constructing a $\mathcal{T}_{\ll}$-word (Definition \ref{def_Tword}) for a new circ-UMFF, making use of Theorem \ref{thm-newcircUMFF}.
We therefore consider a concept analogous to Lyndon and $V$-words based on alternating lexorder:  



\begin{defn} \cite{R06}
\label{def_Galois}
A \itbf{Galois word} is a nonempty primitive string that is minimum in alternating lexorder over its conjugacy class.
\end{defn}

We observe that every element of $\Sigma$ is a Galois word. Thus the set of Galois words forms an FF --- see Section \ref{sect-ummf}.
Since the rotations of primitive strings are distinct, therefore the Galois word of each conjugacy class is uniquely defined. 
Examples of Galois words are: $ab$, $aba$, $abb$, $abba$, $ababa$, $ababaa$, $ababba$. Observe that these words are not necessarily border-free and can even be palindromic. Furthermore, unlike Lyndon words and $V$-words, Galois words don't exhibit the ordered shuffle property \cite{CM-92,DDS13} where interleaving characters/$V$-letters generates a string in the given circ-UMFF; for example, interleaving the Galois words $aba$ and $abb$, where $aba \prec_{alt} abb$,  yields $aabbab$, which is not a Galois word --- instead the conjugate $abaabb$ is Galois. If we try to apply Lemma \ref{lem-xyz} to the Galois words $\s{xy} = ababa$ and $\s{yz} = ab$ with $\s{y} = a$, we get $\s{xyz} = ababab$, a repetition, while Galois words are by definition primitive. Hence 
Galois words do not form an UMFF, nor therefore a circ-UMFF.
Since Galois words can be bordered, this observation is consistent with Theorem \ref{thm-newcircUMFF}, which allows a circ-UMFF only for classes of border-free words. 

A related 
observation in \cite{DBLP:journals/tcs/DolceRR19} shows that, although Theorem \ref{thm-Duval} implies that the unique maximal Lyndon factorization
of a word has a least number of nonincreasing Lyndon factors, this is not necessarily the case for Galois words:

{\begin{examp} [Example 44 in \cite{DBLP:journals/tcs/DolceRR19}]
\label{ex-Galois}
Let \s{w} be the repetition $ababab$, hence not a Galois word. Then $\s{w} = (ab)(ab)(ab)$ is a nonincreasing factorization into Galois words. However, \s{w} also admits a shorter (and increasing) factorization into Galois words, namely $\s{w} = (ababa)(b)$.
\end{examp}
In this example, for $\s{w} = (ababa)(b)$ we have $ababa \prec_{alt} b$, although, since $ababab \notin \mathcal{W}$, $ababa$ is not less than $b$ in $\mathcal{W}$-order 
(Definition \ref{def-W-order}) --- that is, $ababab$ is not a Galois word. 
Note also that $\s{w} = (ab)(ab)(ab)$ is the unique Lyndon factorization of \s{w}.}

{Still, Galois words have applications. The Alternating Burrows-Wheeler Transform (ABWT) is analogous to the BWT and applies alternating lexorder \cite{DBLP:journals/ejc/GesselRR12}. An algorithmic perspective on the ABWT is given in \cite{DBLP:journals/tcs/GiancarloMRRS20}, where a linear time and space algorithm, \texttt{FINDGALOISROTATION}, determines for each primitive string its unique cyclic rotation that is a Galois
word, which in turn supports computing the ABWT in linear time. Furthermore, practical uses of the ABWT in settings such as data compression and
compressed data structures are demonstrated.}

Clearly, like lexorder string comparison, alternating lexorder comparison of two strings can be achieved in linear time and space. It follows that efficient suffix array construction methods based on lexorder, such as \cite{KA03}, can be readily modified to apply alternating lexorder. In the alternating lexorder suffix array of a Galois word with border \s{b}, given that a border of a Galois word must have odd length 
\cite[Lemma 7.3]{DBLP:journals/tcs/GiancarloMRRS20}, then by Definition~\ref{def_alt_lexorder}(2), a suffix starting with the prefix \s{b} will precede
the suffix \s{b}.

We now go on to explore new combinatorial properties of Galois words, starting with the Galois equivalent of the fundamental Lyndon result, Theorem \ref{thm-Duval} --- albeit with the caveat of possible repetitions, previously shown by the example $ababa \prec_{alt} b$.  We let $\mathcal{G}$ denote the set of Galois words on a given alphabet. 

\begin{lemm}(\cite[Proposition 7.4]{DBLP:journals/tcs/GiancarloMRRS20})\label{lem_Galois-suffixes}
A Galois word $\s{w} \in \mathcal{G}$ is smaller than any of its proper suffixes with respect to $\prec_{alt}$ order.
\end{lemm}

\begin{lemm}
\label{lem_Galois-order1}
Let $\s{u},\s{v} \in \mathcal{G}$, where $\s{u}$ is not a proper prefix of $\s{v}$. Then, $\s{u} \prec_{alt} \s{v}$ if and only if\\
    (1) $\s{u} \prec_{alt} \s{vx}, \forall \s{x} \in \Sigma^+$; \\
     (2) $\s{ux} \prec_{alt} \s{v}, \forall \s{x} \in \Sigma^+$; \\
    (3) $\s{ux} \prec_{alt} \s{vx}, \forall \s{x} \in \Sigma^+$; \\
    (4) $\s{u}^{k+1} \prec_{alt} \s{u}^{k}\s{v}$, for $k \geq 0$.
\end{lemm}
\begin{proof} 
{\bf NECESSITY.} Suppose $\s{u} \prec_{alt} \s{v}$. Since $\s{u}$ is not a proper prefix of $\s{v}$, therefore $\forall \s{x} \in \Sigma^*$,  $\s{u}$ is not a proper prefix of $\s{vx}$, and $\s{ux}$ is not a proper prefix of $\s{v}$ and $\s{vx}$ --- and $\s{u}^{k+1}$ is not a prefix of $\s{u}^{k}\s{v}$. Moreover, since $\s{u} \prec_{alt} \s{v}$ and $\s{u}$ is not a proper prefix of $\s{v}$, there is a smallest index $i$ such that $u_i \neq v_i$. Consider a corresponding relation $R$ ($<$ or $>$ for odd or even index $i$, respectively) --- then since $u_i \neq v_i$, it follows that $u_i~R~v_i$ implies $(u)_i~R~(vx)_i$, $(ux)_i~R~(v)_i$, $(ux)_i~R~(vx)_i$ and $(u^{k+1})_i~R~    (u^{k}v)_i$. Therefore, by Definition~\ref{def_alt_lexorder}(1) cases (1)--(4) hold. 

{\bf SUFFICIENCY.} Suppose cases (1)--(4) hold. Since $\s{u}$ is not a proper prefix of $\s{v}$, by the same reasoning as above, we get $(u)_i~R~(vx)_i$, $(ux)_i~R~(v)_i$, $(ux)_i~R~(vx)_i$ and $(u^{k+1})_i~R~(u^{k}v)_i$ which implies that $u_i~R~v_i$. Therefore, by Definition~\ref{def_alt_lexorder}(1), $\s{u} \prec_{alt} \s{v}$. 
\qed
 \end{proof}

\begin{lemm}
\label{lem_Galois-order2}
Let $\s{u},\s{v} \in \mathcal{G}$. If $\s{u} \prec_{alt} \s{v}$ and $\s{u}$ is a proper prefix of $\s{v}$, then\\
    (1) $\forall \s{x} \in \Sigma^+, \s{u} \prec_{alt} \s{vx}$; \\
    (2) $\forall k \geq 0, \s{u}^{k+1} \prec_{alt} \s{u}^{k}\s{v}$.
     
\end{lemm}
\begin{proof}
Suppose $\s{u} \prec_{alt} \s{v}$ and $\s{u}$ is a proper prefix of $\s{v}$. By Definition~\ref{def_alt_lexorder}(2), $|\s{u}|$ must be even. We write $\s{v} = \s{u}^r\s{v}'$, such that $r\geq 1$ and $\s{v}'$ is the largest suffix of $\s{v}$ that does not contain $\s{u}$ as its prefix. Then, $\s{u^kv} = \s{u}^{k+r}\s{v}'$. Clearly, $\s{u}$ is a prefix of $\s{vx}$ and $\s{u}^k\s{v}$. By hypothesis $|\s{u}|$ is even. Hence, by Definition~\ref{def_alt_lexorder}(2), cases (1) and (2) hold.
\qed
 \end{proof}
 
\begin{thrm}
\label{thrm_Galois-order}
Suppose $\s{u}, \s{v} \in 
\mathcal{G}$. If $\s{u}\s{v}$ is primitive,  then $\s{u} \prec_{alt} \s{v}$ if and only if $\s{u}\s{v} \in \mathcal{G}$.
\end{thrm}
\begin{proof}
{\bf NECESSITY.} Suppose that $\s{u} \prec_{alt} \s{v}$. Since $\s{uv}$ is primitive, there exists a rotation of $\s{uv}$ that is least in alternating lexorder ($\prec_{alt}$). Let $\s{u} = \s{u}_p\s{u}_s$ and $\s{v}=\s{v}_p\s{v}_s$, where $\s{u}_p, \s{u}_s, \s{v}_p, \s{v}_s \neq \varepsilon$. Then, we need to show
\begin{enumerate}
    \item[(1)] $\s{uv} \prec_{alt} \s{u}_s\s{v}\s{u}_p$;
    \item[(2)] $\s{uv} \prec_{alt} \s{vu}$;
    \item[(3)] $\s{uv} \prec_{alt} \s{v}_s\s{u}\s{v}_p$.
\end{enumerate}
{\bf (A)} Suppose $\s{u}$ is not a proper prefix of $\s{v}$. By Lemma~\ref{lem_Galois-suffixes}, we have $\s{u} \prec_{alt} \s{u}_s$, and by Lemma~\ref{lem_Galois-order1}(3) and (1), we have $\s{uv} \prec_{alt} \s{u}_s\s{v}$ and $\s{uv} \prec_{alt} \s{u}_s\s{v}\s{u}_p$, respectively. Therefore case (1) holds. From the hypothesis and by Lemma~\ref{lem_Galois-suffixes} we have $\s{u} \prec_{alt} \s{v}_s$. By Lemma~\ref{lem_Galois-order1}(2) and (1), we have $\s{uv} \prec_{alt} \s{v}_s$ and $\s{uv} \prec_{alt} \s{v}_s\s{u}\s{v}_p$, respectively, and so case (3) holds. Finally, from the hypothesis we have $\s{u} \prec_{alt} \s{v}$. By the application of Lemma~\ref{lem_Galois-order1}(1) and (2), we have $\s{u} \prec_{alt} \s{vu}$ and $\s{uv} \prec_{alt} \s{vu}$, respectively. Thus case (2) holds. 

{\bf (B)} Suppose $\s{u} \prec_{alt} \s{v}$, and $\s{u}$ is a proper prefix of $\s{v}$. 
By Lemma~\ref{lem_Galois-suffixes}, we have $\s{u} \prec_{alt} \s{u}_s$. By Lemma~\ref{lem_Galois-order1}(1) and (2) we get $\s{u} \prec_{alt} \s{u}_s\s{v}\s{u}_p$ and $\s{u}\s{v} \prec_{alt} \s{u}_s\s{v}\s{u}_p$, respectively. Thus case (1) holds.  
By hypothesis and Lemma~\ref{lem_Galois-suffixes}, we get $\s{u} \prec_{alt} \s{v}_s$, and by Lemma~\ref{lem_Galois-order1}(1) and (2) we get $\s{u} \prec_{alt} \s{v}_s\s{u}\s{v}_p$ and $\s{u}\s{v} \prec_{alt} \s{v}_s\s{u}\s{v}_p$, respectively. Hence case (2) holds.  
We write $\s{v} = \s{u}^r\s{v}'$, such that $r\geq 1$ and $\s{v}'$ is the largest suffix of $\s{v}$ that does not contain $\s{u}$ as its prefix. Since $\s{uv}$ and $\s{vu}$ are of the same length, neither can be a proper prefix of the other. By the hypothesis and Lemma~\ref{lem_Galois-suffixes} $\s{u} \prec_{alt} \s{v}'$ and $\s{u}$ is not a proper prefix of $\s{v}'$.  Then $\s{uv} = \s{u}^{k+1}\s{v}'$ and $\s{vu} = \s{u}^k\s{v}'\s{u}$. By Lemma~\ref{lem_Galois-order1}(4), we get $\s{u}^{k+1} \prec_{alt} \s{u}^k\s{v}'$. By further application of Lemma~\ref{lem_Galois-order1}(1) and (2), we get $\s{u}^{k+1} \prec_{alt} \s{u}^k\s{v}'\s{u}$ and $\s{u}^{k+1}\s{v}' \prec_{alt} \s{u}^k\s{v}'\s{u}$, respectively. Thus $\s{uv} \prec_{alt} \s{vu}$ and case (3) holds. \\
{\bf SUFFICIENCY.} Suppose that primitive $\s{uv} \in \mathcal{G}$. Then $\s{u} \neq \s{v}$, and $\s{u}\s{v}$ is strictly least in its conjugacy class in $\prec_{alt}$ order. Suppose $\s{v} \prec_{alt} \s{u}$. Then, by the proof of Necessity, we find that $\s{vu}$ is a Galois word --- a contradiction. Since $\s{u} \neq \s{v}$ and $\s{v} \not \prec_{alt} \s{u}$, Theorem~\ref{thm-altorder} implies that $\s{u} \prec_{alt} \s{v}$.

Thus the theorem is proved.
\qed
\end{proof}

We formalize the notion of concatenation order $\mathcal{W}$ (Definition \ref{def-W-order}) ``aligning'' with   
$\mathcal{T}$-order 
(Definition \ref{def_Torder}):
\begin{defn}
\label{def_UMFF-align}
Given a circ-UMFF $\mathcal{W}$, a concatenation order $<_{\mathcal {W}}$  
and a $\mathcal{T}$-order
with order relation $\ll$,
we say that {\bf $\mathcal{W}$ is $\mathcal{T}$-order aligned}
if, for given strings $\s{u}, \s{v} \in \Sigma^*$, $\s{u} <_\mathcal{W} \s{v}$ and $\s{u} \ll \s{v}$.
\end{defn}
The classic example is the lexorder alignment of Lyndon words, as expressed in Theorem \ref{thm-Duval}. On the other hand, this alignment doesn't hold for co-Lyndon words (the co-lexorder analog of Lyndon words where co-lexorder is lexorder of reversed strings): 
$ba$ is less than $ca$ in co-lexorder, while $baca$ is not a co-Lyndon word --- instead, it is 
the conjugate $caba$. A further example of non-alignment arises with $V$-order --- see Lemma 3.16 in \cite{DDS13}. We have seen that Galois words do not form a circ-UMFF and, furthermore, Theorem \ref{thrm_Galois-order} shows that alternating lexorder $\prec_{alt}$ is not in general aligned with Galois concatenation.\\

We now show that the \itbf{set of border-free Galois words} over any alphabet, denoted $\mathcal{G}^{bf}$, yields a unique maximal factorization: 
\begin{lemm}
\label{lem_Galois-bf}
$\mathcal{G}^{bf}$ forms an UMFF.
\end{lemm}
\begin{proof}
We will show that
the $\mathcal{G}^{bf}$ satisfies the $\mathbf{xyz}$ Lemma \ref{lem-xyz}. 
Thus we suppose that $\s{xy},\s{yz} \in \mathcal{G}^{bf}$ for some nonempty \s{y}; it is then required to show that $\s{xyz} \in \mathcal{G}^{bf}$. We may assume also that both $\s{x}$ and $\s{z}$ are nonempty, for otherwise by Lemma \ref{lem-xyz} the claim holds trivially.

We start by showing that $\s{xyz}$ is least in $\prec_{alt}$ order amongst its conjugates. 
Write $\s{xyz} = x_1 \ldots x_r y_1 \ldots y_s  z_1 \ldots z_t$, $r,s,t \ge 1$, and consider ordering the conjugates of $\s{xyz}$ in $\prec_{alt}$ order (Definition~\ref{def_alt_lexorder}). Let $\mathcal{C}^{\prec_{alt}}$ denote the conjugacy class of \s{xyz}, where the conjugates are ordered according to ${\prec_{alt}}$.
Since a Galois word is less than any of its proper suffixes in $\prec_{alt}$ order 
\cite[Proposition 7.4]{DBLP:journals/tcs/GiancarloMRRS20}, this property therefore holds for $\s{xy}$ and $\s{yz}$. For the ordering of $\mathcal{C}^{\prec_{alt}}$, consider a conjugate \s{c} of $\s{xyz}$ starting with a nonempty suffix \s{s} of $x_2 \ldots x_r y_1 \ldots y_s$. Since $\s{xy}$ is assumed to be border-free, $\s{xy}[1..|\s{s}|] \neq \s{s}$ and thus the comparison between $\s{xyz}$ and \s{c} is determined based on Definition \ref{def_alt_lexorder}(1). Therefore
the conjugate of $\s{xyz}$ starting with prefix $\s{xy}$, namely $\s{xyz}$, comes before any conjugate in $\mathcal{C}^{\prec_{alt}}$ starting with a suffix of $x_2 \ldots x_r y_1 \ldots y_s$
and in particular starting with \s{y}. 
That is, if $\s{xy}= \s{p}\s{s}$ for nonempty $\s{p}$, $\s{s}$, then $\s{xy} \prec_{alt} \s{s}$ implies that $\s{xyz} \prec_{alt} \s{s} \prec_{alt} \s{s} \s{z} \s{p}$.
Likewise, the conjugate of $\s{xyz}$ starting with prefix $\s{yz}$, namely \s{yzx}, comes before any conjugate in $\mathcal{C}^{\prec_{alt}}$ starting with a suffix of  $y_2 \ldots y_s z_1 \ldots z_t$, and as before we can assume this is determined using Definition \ref{def_alt_lexorder}(1).  
We conclude that $\s{xyz}$ precedes all its conjugates in $\mathcal{C}^{\prec_{alt}}$, as required.

Suppose now that $\s{xyz}$ is bordered with border $\s{b}$ so that $\s{xyz} = \s{b}\s{u}\s{b}$ for $\s{u} \in \Sigma^*$. Then \s{b} is both a prefix of \s{xy} and a suffix of \s{yz}. Since distinct $\s{xy}, \s{yz} \in  \mathcal{G}^{bf}$, both are border-free, and so \s{xy} cannot have suffix \s{b} nor can \s{yz} have prefix \s{b}. We can then write $\s{xyz}= \s{b}\s{u}\s{b} = b_1b_2 \ldots b_r   u_1u_2 \ldots u_s   b_1b_2 \ldots b_r$, $r,s \ge 1$. If we suppose that $|\s{xy}| > r+s$, then \s{xy} is bordered; similarly, if $|\s{yz}| > s+r$, then \s{yz} is bordered. 
So $\s{y} = u_i \ldots u_j$, $1 \le i$, $j \le s$. 
The smallest and first conjugate in $\mathcal{C}^{\prec_{alt}}$ is $\s{xyz} = \s{x} u_i \ldots u_j \s{z} = \s{b} u_1 \ldots u_s \s{b}$, and since the ordering of $\mathcal{C}^{\prec_{alt}}$ applies 
(Definition \ref{def_alt_lexorder}(1)), common prefixes occur concurrently in $\mathcal{C}^{\prec_{alt}}$; in particular, if there are $n_b$ occurrences of \s{b} in \s{xyz}, they will occur as prefixes of the first $n_b$ conjugates in $\mathcal{C}^{\prec_{alt}}$. 
One of the conjugates in $\mathcal{C}^{\prec_{alt}}$ is $\s{yzx}$, where $\s{yzx} = u_i \ldots u_s \s{bx}$ and $\s{xyz} \prec_{alt} \s{yzx}$. However, since \s{b} is a prefix of the first conjugate \s{xyz}, this contradicts that $\s{yz} = u_i \ldots u_s \s{b}$ is less than its proper suffix \s{b} in ${\prec_{alt}}$ order. We conclude that \s{xyz} is border-free, and therefore primitive, which completes the proof.
\qed
\end{proof}
We illustrate Lemma \ref{lem_Galois-bf}:
\begin{examp}
\label{examp-Galoisbf}
 Let $\s{xy, yz} \in \mathcal{G}^{bf}$, where $\s{xy}= abababbb$ and $\s{yz}= ababbbbb$ and $\s{y}= ababbb$ then, applying Lemma \ref{lem-xyz}, $\s{xyz} = abababbbbb$ with $\s{xyz} \in \mathcal{G}^{bf}$.
\end{examp}

\subsection{Binary Border-free Galois words}\label{ssec-Galois-bbf}
We next consider the structure of binary border-free Galois words, denoted $\mathcal{G}^{bbf}$, where we assume that 
$\Sigma = \{a, b\}$ with $a<b$. The first few such words are $a$, $b$, $ab$, $abb$.

\begin{lemm}
\label{claim_Galois-bbf}
Suppose $\s{w} \in \mathcal{G}^{bbf}$ over $\Sigma = \{a, b\}$, where $a < b$, with $n = |\s{w}| \ge 3$. Then $\s{w}$ has prefix $ab$ and suffix $bb$.
\end{lemm}
\begin{proof}
By definition \s{w} 
is primitive and so contains 
both letters in $\Sigma$.
Also by definition, \s{w} is minimum over its rotations,
and so starts with $a$; since it is border-free, it must therefore end with $b$. 
The claim clearly holds for $\s{w} = ab^k$, $k \ge 2$, 
as all non-trivial rotations of $\s{w}$ start with $b$ making $\s{w}$ least in $\prec_{alt}$ order. 
So suppose that $\s{w} = a\s{v}b$ where $\s{v} \in \Sigma^+$:  
\begin{itemize}
\item If $\s{w}[n-1] = a$, then the suffix is $ab$. Let $\s{u}$ denote the conjugate of $\s{w}$ such that 
$\s{u} = \s{w}[n-1]\s{w}[n]\s{w}[1..n-2]$. Since $\s{w}$ is least in $\prec_{alt}$ order, and the comparison of $\s{w}[2]$ and $\s{u}[2]$ is based on the relation $\ge$, therefore $\s{w}[2]$ cannot be $a$, thus
yielding prefix $ab$ and so a bordered word.  Hence $\s{w}[n-1] = b$ and the suffix is $bb$, as required.
\item  
Since $\s{w} \neq ab^k$, $k \ge 3$, therefore $\s{v}$ 
contains an $a$. Let $a_i$ be the $a$ in $\s{v}$ with largest $i$, so that $1< i <n-1$, and since the string $\s{w}$ must end with $bb$ we know that the $(i+1)^{\rm th}$ letter is $b$. Let 
$\s{u'} = \s{w}[i]\s{w}[i+1]..\s{w}[n]\s{w}[1 .. i-1]$. Since  $\s{w}$ is least in $\prec_{alt}$ order, therefore as above $\s{w}[2]$ cannot be $a$, 
thus yielding prefix $ab$.  
\end{itemize}
We conclude that $\s{w}$ has prefix $ab$ and suffix $bb$, as required.
\qed
\end{proof}

We next consider the general form of a binary border-free Galois word \s{w}, $|\s{w}| >3$. By Lemma \ref{claim_Galois-bbf},  if $\s{w} \in \mathcal{G}^{bbf}$, \s{w} begins with $ab$. Trivially if  $\s{w} = ab^h$, $h>1$, then $\s{w} \in \mathcal{G}^{bbf}$. So suppose that $\s{w} \neq ab^h$, thus containing at least two $a$'s --- since any conjugate of $\s{w}$ starting with $b$ is consistent with $\s{w} \in \mathcal{G}^{bbf}$, we restrict analysis to conjugates starting with $a$.  

Consider a run in $\s{w}$ of the form $a^k$, $k>1$. Then since $\s{w}$ has prefix $ab$, any conjugate of $\s{w}$ starting with $a^j$, $2 \le j \le k$, is consistent with $\s{w} \in \mathcal{G}^{bbf}$. So we consider conjugates $\s{c_{ab}}$ of \s{w} starting with $ab$ and note the index $d$ of the first difference between \s{w} and $\s{c_{ab}}$ --- since \s{w} is primitive, \s{w} and $\s{c_{ab}}$ are distinct.



We say a substring \s{u} of $\s{w}$ is an \itbf{ab-Galois word} if \s{u} has prefix $ab$ and contains no other occurrence of $ab$. Therefore \s{u} has the form $a b^e a^f$, $e \ge 1$, $f \ge 0$. Then \s{u} is a Galois word (not necessarily border-free): clearly \s{u} is primitive, and it is less in alternating lexorder $\prec_{alt}$ than any conjugate of \s{u} starting with $b$ and likewise less than any conjugate starting with $aa$, if such exists.\\ 

\begin{lemm}
\label{lem_ab-pattern}
Given a primitive binary string
$\s{w}$,  
let $\mathcal{F}$ be the factorization of \s{w} into $ab$-Galois words $\s{u}_1,\s{u}_2\cdots \s{u}_t$, such that $\mathcal{F} = (\s{u}_1) (\s{u}_2) \cdots (\s{u}_t)$, $t \ge 1$. 
If $|\s{u}_1| = |\s{u}_2| = \cdots = |\s{u}_t|$, and $\s{w}$ is a Lyndon word under $\mathcal{T}_{lex}$-order (Definition \ref{Torder}) with  $\prec_{alt}$ order, then $\mathcal{F} \in \mathcal{G}^{bbf}$.
\end{lemm}

\begin{proof}
Trivially this holds for any conjugates starting with $aa$ or $b$. For conjugates starting with $ab$, we first observe that Theorem  \ref{thm-altorder} shows that alternating lexorder $\prec_{alt}$ is a candidate for $\mathcal{T}$-order (Definition \ref{def_Torder}). Then the result is an immediate consequence of maintaining Lyndon properties for conjugates starting with $ab$-Galois words when generalising lexorder to lex-extension order. In particular, \s{w} is border-free.
\qed
\end{proof}

For the next more general result, we require a refinement of Definition \ref{def_alt_lexorder}(2). We define modified alternating lexorder $\prec_{modalt}$ as follows: 



\begin{defn}[Modified Alternating lexorder $\prec_{modalt}$] \label{def_mod_alt}
Let $\s{x}, \s{y} \in \Sigma^+$ be distinct nonempty strings, where $\s{x}$ is a proper prefix of $\s{y}$. Let $S$ be the largest set of strings in $\Sigma^+$ such that for every $\s{z} \in S$, $\s{x}\s{z} \prec_{alt} \s{y}$, and neither $\s{x}\s{z}$ is a prefix of $\s{y}$ nor is $\s{y}$ a prefix of $\s{x}\s{z}$. Then, if $S \neq \emptyset$, $\s{x} \prec_{modalt} \s{y}$; otherwise, $\s{y} \prec_{modalt} \s{x}$. 
\end{defn}

\begin{lemm}\label{lem-modalt}
   Modified alternating lexorder $\prec_{modalt}$ 
   is a total order over $\Sigma^*$.
\end{lemm}

\begin{proof} 
Consider distinct strings $\s{x}, \s{y}, \s{t} \in \Sigma^+$. Assume that $\s{x} \prec_{modalt} \s{y}$ and $\s{y} \prec_{modalt} \s{t}$. We need to show that $\s{x} \prec_{modalt} \s{t}$.

By assumption and Definition~\ref{def_mod_alt}, $\s{x}$ is a nonempty proper prefix of $\s{y}$ and $\s{y}$ is a nonempty proper prefix of $\s{t}$. Hence we write $\s{y}=\s{xy}'$, where $\s{y}' \neq \varepsilon$, and $\s{t} = \s{yt}'$, where $\s{t}' \neq \varepsilon$. Then, by substitution, we get $\s{t}= \s{x}\s{y}'\s{t}'$. Also, by assumption and Definition~\ref{def_mod_alt}, we may suppose $S, S' \subset \Sigma^+$ are nonempty sets satisfying the following conditions:
\begin{enumerate}
    \item $\s{xz} \prec_{alt} \s{y}$, where for every $\s{z} \in S$, neither $\s{xz}$ is a prefix of $\s{y}$ nor is $\s{y}$ a prefix of $\s{xz}$.
    \item $\s{yz}' \prec_{alt} \s{t}$, where for every $\s{z}' \in S'$, neither $\s{yz}'$ is a prefix of $\s{t}$ nor is $\s{t}$ a prefix of $\s{yz}'$.
\end{enumerate}

Since $\s{xz}$ is not a proper prefix of $\s{y}$, by Lemma~\ref{lem_Galois-order1}(1) and (2), we get $\s{xz} \prec_{alt} \s{yz}'$ and $\s{xzz}'' \prec_{alt} \s{yz}'$, where $\s{z}' \in S', \s{z}'' \in \Sigma^+$, respectively. Then, by Theorem~\ref{thm-altorder}, we get $\s{xzz}'' \prec_{alt} \s{t}$. Since $\s{xz}$ is not a prefix of $\s{y}$ nor $\s{y}$ a prefix of $\s{xz}$, and since $\s{y}$ is a prefix of $\s{t}$, we conclude that $\s{xz}$ and $\s{t}$ are not prefixes of each other, nor are $\s{xzz}''$ and \s{t} prefixes of each other. Let $S''$ denote the set containing all strings of the form $\s{zz}''$. Clearly, $S'' \neq \emptyset$.  Therefore, by Definition~\ref{def_mod_alt}, we conclude that $\s{x} \prec_{modalt} \s{t}$.
This completes the proof.\qed
\end{proof}

We will also apply the concept of a Hybrid Lyndon word, whereby a string \s{x} in $V$-form with $\s{x_0} = \s{\varepsilon}$ is a Hybrid Lyndon if and only if it is Lyndon under lexicographic extension \cite[Definition 3.12]{DDS13}.

\begin{thrm}
\label{thm_ab-pattern}
Given a primitive binary string \s{w}, with $|\s{w}| > 3$, let $\mathcal{F}$ be the factorization of \s{w} into $ab$-Galois words $\s{u}_1,\s{u}_2\cdots \s{u}_t$, such that $\mathcal{F} = (\s{u}_1) (\s{u}_2) \cdots (\s{u}_t)$, $t \ge 1$. 
Then $\s{w} \in \mathcal{G}^{bbf}$ if and only if $\mathcal{F}$ forms a Hybrid Lyndon word using $\mathcal{T}_{lex}$-order (Definition \ref{Torder}) with $\prec_{modalt}$ order.
\end{thrm}

\begin{proof}

NECESSITY. Suppose $\s{w} \in \mathcal{G}^{bbf}$ with $|\s{w}| > 3$. From Claim \ref{claim_Galois-bbf}, \s{w} starts with $ab$, namely an $ab$-Galois word, which is consistent with \s{w} being less in $\prec_{alt}$ order than any conjugate \s{c} of \s{w} starting $aa$ or $b$.\\ 

Hence we restrict our analysis to conjugates starting with $ab$, that is $ab$-Galois words (factors of $\mathcal{F}$). Consider the comparison of $\s{w} = (\s{u}_1) (\s{u}_2) \cdots (\s{u}_t) = \s{w}[1..t]$ and the $i^{th}$ substring conjugate
$\s{c_i}$ of \s{w} where both have prefix $ab$, so that  $\s{c_i} = (\s{u_{i+1}}) (\s{u_t}) (\s{u_1}) \cdots (\s{u_i}) = \s{c_i}[1..t]$. Since \s{w} is primitive, \s{w} and \s{c_i} are distinct, and as $\s{w} \in \mathcal{G}^{bbf}$ then $\s{w} \prec_{alt} \s{c_i}$ which is decided according to Definition \ref{def_alt_lexorder}. Let $d$ be the index of the first distinct $ab$-Galois words  between \s{w} and $\s{c_i}$, that is 
$\s{w}[1..d-1] = \s{c_i}[1..d-1]$, and let $h = min \{|\s{w}[d]|, |\s{c_i}[d]|\}$.\\

Then according to Definition \ref{def_alt_lexorder}, there are two cases to consider:\\

\noindent Definition \ref{def_alt_lexorder}(1). In this case the two factors differ at an index $q \le h$, that is $\s{w}[g+q] \neq \s{c_i}[g+q]$, where $g = \sum_{j=1}^{d-1}|\s{u_j}|$. Since $\s{w} \prec_{alt} \s{c_i}$, we have $\s{w}[g+q] ~R~ \s{c_i}[g+q]$, where $R = <$ if $g+q$ is odd and $R = >$ if $g+q$ is even. This holds for any conjugate satisfying Definition \ref{def_alt_lexorder}(1) and thus this case is consistent with $\mathcal{F}$ forming a Hybrid Lyndon word using $\mathcal{T}_{lex}$-order (Definition \ref{Torder}) with $\prec_{modalt}$ order (which is in fact regular $\prec_{alt}$ order here).\\

\noindent Definition \ref{def_alt_lexorder}(2) (Notation as in part (1).) In this case, with $\s{w}[d] \neq \s{c_i}[d]$, one factor is a proper prefix of the other factor. Since $\s{w} \in \mathcal{G}^{bbf}$ and $\s{w} \prec_{alt} \s{c_i}$, if $\s{w}[d]$ is a proper prefix of $\s{c_i}[d]$, then $|\s{u_d}| = h < |\s{c_i}[d]|$, whereby $d<t$. Then applying $\prec_{modalt}$ order, 
$\s{w}[d+1]$ has prefix $ab$ while, by definition of an $ab$-Galois word, 
$\s{c_i}[d]$ has only one occurance of $ab$ which occurs at its prefix. So the comparison which determines $\s{w} \prec_{alt} \s{c_i}$ is between $\s{w}[d+1]$ and $\s{c_i}[d]$, in particular between an $a$ and $b$ or between $ab$ and $aa$ with the appropriate relations $(< / >)$. Similarly, if $\s{c_i}[d]$ is a proper prefix of $\s{w}[d]$, then $\prec_{modalt}$ order applies as appropriate to ensure $\s{w} \prec_{alt} \s{c_i}$.
This ordering holds for any conjugate satisfying Definition \ref{def_alt_lexorder}(2) and thus this case is consistent with $\mathcal{F}$ forming a Hybrid Lyndon word using $\mathcal{T}_{lex}$-order (Definition \ref{Torder}) with $\prec_{modalt}$ order.\\ 

SUFFICIENCY. Suppose the Hybrid Lyndon condition in the statement holds for a primitive binary string \s{w}. Clearly, \s{w} is less than any conjugate \s{c} of \s{w} starting $aa$ or $b$ with respect to $\prec_{alt}$ order. So consider any conjugate \s{c} starting with an  $ab$-Galois word. Accordingly, the Hybrid Lyndon constructed from $\mathcal{T}_{lex}$-order (Definition \ref{Torder}) with $\prec_{modalt}$ order guarantees that $\s{w} \prec_{alt} \s{c}$ for all conjugates starting with an $ab$-Galois word.  Hence all forms of conjugates are considered. Furthermore, from Lyndon properties, \s{w} is border-free. Thus, $\s{w} \in \mathcal{G}^{bbf}$.
\qed
\end{proof}

\begin{examp}
\label{ex-G_bbf}    
Let $\s{w} = abbbabbabbbbb$ be a binary primitive string, so that $\mathcal{F} = (abbb) (abb) (abbbbb)$ with $t = 3$. Then, for instance, $\s{w} = (abbb) (abb) (abbbbb) \prec_{modalt} (abb) (abbbbb) (abbb)$ since $(abbb) \prec_{modalt} (abb) (abbbbb)$, 
and since   $(abbb)(abb) \prec_{modalt} (abbbbb)$, $\s{w} = (abbb) (abb) (abbbbb) \prec_{modalt} (abbbbb) (abbb) (abb)$.
\end{examp}

In considering whether the set $\mathcal{G}^{bbf}$ forms an UMFF, as in Lemma \ref{lem_y-empty}, let $\s{u} = abbbbb$ and $\s{v} = ababbb$, where $\s{u,v} \in \mathcal{G}^{bbf}$. But observe that $\s{uv}$ is bordered with border $abbb$, indicating that $\mathcal{G}^{bbf}$ may not form an UMFF. On the other hand, applying Lemma \ref{lem-xyz} with $\s{xy} = \s{v}$ and $\s{yz} = \s{u}$, where $\s{y} = abbb$, yields $\s{xyz} = ababbbbb$, which does belong to $\mathcal{G}^{bbf}$ indicating $\mathcal{G}^{bbf}$ may form an UMFF.

\section{UMFF to circ-UMFF construction}
\label{sec_UMFF2cUMFF}

In this final section we propose a method for circ-UMFF construction that is based on 
Theorem 4.1 in \cite{DD08} --- which states that any binary border-free UMFF can be enlarged to a binary circ-UMFF. 
Our algorithm takes as input 
a binary border-free UMFF $\mathcal{W}$ of size $n$, where the words are given in increasing length, and outputs an enlargement of $\mathcal{W}$ to a finite circ-UMFF which is necessarily border-free. We assume $|\mathcal{W}| > 2$; otherwise, the UMFF would contain only the alphabet.  Thus we may choose a known circ-UMFF such as binary Lyndon words or binary $V$-words. The algorithm is in 3 stages:
\begin{itemize}
\item input the UMFF and record where word lengths change; \item generate all strings in $\Sigma^*$ up to length $2^{|\s{w_n}|+1} - 2$, where $|\s{w_n}|$ is the length of the longest word in $\mathcal{W}$; 
\item generate the primitive border-free circ-UMFF words using $\s{w_3}=[w_1 \ldots w_m]$; that is, $[w_m w_1]$ is added to the circ-UMFF (assumed necessarily).
\end{itemize}The subset of $\Sigma^*$ is computed in \texttt{procedure Create-$\Sigma^*$} (Algorithm~\ref{alg-Kleene*}) and all variables are assumed to be global.

\begin{algorithm}  
\caption{Generate a finite subset of Kleene star}\label{alg-Kleene*}
{\bf procedure} Create-$\Sigma^*$($\mathcal{W}$)
\begin{algorithmic}

\State $total \gets 2^{|\s{w_n}|+1} - 2$ \Comment{{\it Finite subset of Kleene star generated}}
\State $\Sigma^*[1] \gets \sigma_1$, $\Sigma^*[2] \gets \sigma_2$  \Comment{{\it Array $\Sigma^*$ contains Kleene star}}
\State $s \gets 1, t \gets 2$  \Comment{{\it Indexing start and end of words of same length}}
\State $number \gets t-s+1$
\State $k \gets t+1$ \Comment{{\it $k$ indexes the array $\Sigma^*$}}

\Repeat  \Comment{{\it Generate words in $\Sigma^*$ of maximum length $|\s{w_n}|$}}
\For{$i = s..t$}
\State $\Sigma^*[k] \gets  \sigma_1  \circ \Sigma^*[i]$
\State $\Sigma^*[k+number]  \gets  \sigma_2  \circ \Sigma^*[i]$

\State $k++$
\EndFor

\State $number \gets 2*number$
\State $s \gets t+1$, $t \gets t+number$
\State $k \gets t + 1$

\Until{$k > total$} 
\[\]

\end{algorithmic}
\end{algorithm}

\begin{algorithm}  
\caption{Enlarge a binary border-free UMFF to a circ-UMFF}\label{alg-UMFF2circUMFF}
{\bf procedure} Create-circUMFF($\mathcal{W}$, $n$)
\begin{algorithmic}
\State  $\mathcal{W} \rightarrow UMFF[\s{w_1}, \s{w_2}, \ldots, \s{w_n}]$ 
\Comment{{\it Read UMFF $\mathcal{W}$ into array UMFF}}
\State $\sigma_1 \gets \s{w_1}$, $\sigma_2 \gets \s{w_2}$ 
\Comment{{\it Assign the binary $\mathcal{W}$ alphabet to constants}} 
\State $i \gets 1,  j \gets 2$
\State $Length[i] \gets j, i++, j++$ \Comment{{\it Dynamic array indexes change in word length}}
\Repeat
\While{$|UMFF[j]| = |UMFF[j+1]|$} $j++$
\EndWhile
\State $Length[i] \gets j, i++,  j++$
\Until {$j \ge n$}
\[\] 

\State Create-$\Sigma^*$($\mathcal{W}$)
\Comment{{\it Generate a finite subset of Kleene star}}\\






\State $circUMFF[1] \gets \sigma_1$ 
\Comment{{\it Dynamic array will contain the circUMFF}}
\State $circUMFF[2] \gets \sigma_2$ \Comment{{\it Added the alphabet to the output circUMFF array}}
\State $circUMFF[3] \gets \s{w_3}[1] \circ \s{w_3}[m]$ \Comment{{\it by Lemma \ref{lem-xyz} word can't be $\s{w_3}[m]\s{w_3}[1]$ }}
\State $k \gets 3$; $j \gets 2$ \Comment{{\it $k$ indexes circUMFF array; $j$ indexes Length array}}

\For 
{$i = 7..total$} \Comment{{\it $i$ indexes words in the $\Sigma^*$ array with length $ > 2$}}
\State $word \gets \Sigma^*[i]$

\If
{$word$ is primitive} \Comment{{\it Linear test}}
\State k++
\While{$|UMFF[Length[j]]| < |word|$} {j++}
\EndWhile

\If
 {$word \in UMFF[Length[j-1]+1..Length[j]]$} 
\State {$circUMFF[k]  \gets word$}  \Comment{{\it UMFF $word$ is border-free}}


\Else
\For
{$s = 1 .. k-1$}
\State $t \gets k-s$   \Comment{{\it assume $s \neq t$}}

\State $string \gets circUMFF[s] \circ circUMFF[t]$ \Comment{{\it Theorem \ref{thm-struct}(3)}}

\State $c=0$

\While {$string \neq R_c(word)$ and $c < |word|$} {$c++$}
\EndWhile

\If
{$string = R_c(word)$}
\Comment{{\it Conjugacy class $R$ contains $string$ so apply $\mathbf{xyz}$ closure of Lemma \ref{lem-xyz} with elements in circUMFF}}

\If
{$string ~\mathbf{xyz}$ $circUMFF[\ell], \ell=3..k-1$ border-free and $string \neq circUMFF[\ell]$} 



\State $circUMFF[k] \gets string$
\EndIf
\EndIf 
\EndFor


\EndIf
 

\EndIf

\EndFor

\end{algorithmic}
\end{algorithm}

\begin{thrm}
\label{thm-alg}
Algorithm ~\ref{alg-UMFF2circUMFF} correctly computes a finite circ-UMFF from a finite binary border-free UMFF.
\end{thrm}

\begin{proof}
This follows from the method of the proof of Theorem 4.1 in \cite{DD08} which states that any border-free binary UMFF can be enlarged to a binary circ-UMFF. All words in Kleene star up to a specified length are considered for the new circ-UMFF and each candidate word is checked to be primitive --- all as required by Definition~\ref{def-subcirc}.
\qed    
\end{proof}

The rationale of Algorithm~\ref{alg-UMFF2circUMFF} is as follows.  
 Given, for instance, the border-free UMFF $\mathcal{G}^{bf}$ over $\Sigma \{a,b\}$, we consider extending a subset of $\mathcal{G}^{bf}$ to a new circ-UMFF which is necessarily border-free. 
 For example, suppose $abababbb \in \mathcal{G}^{bf}$, and consider the primitive string $\s{w} = ababba$, where $\s{w} \in \{\mathcal{G} \setminus \mathcal{G}^{bf}\}$; that is, \s{w} is a bordered Galois word.  According to Definition \ref{def-subcirc}, we need to choose one border-free conjugate from the conjugacy class of $\s{w}$ --- Lyndon words show that a border-free conjugate of a primitive string always exists, so there will be at least one border-free conjugate to choose from. For example, applying Theorem \ref{thm-struct}(5), we select the border-free conjugate $aababb$ of \s{w}, where $a, ababb \in \mathcal{G}^{bf}$, and note that the bordered word $ababba$ is the Galois conjugate of \s{w} while $aababb$ happens to be the Lyndon conjugate of \s{w}. Note further that, with reference to Lemma \ref{lem_y-empty}, $abababbb$ and $aababb$ are mutually border-free and so we cannot generate another word for the new circ-UMFF from them by applying Lemma \ref{lem-xyz}.

Algorithm \ref{alg-UMFF2circUMFF} generates a finite circ-UMFF but can be easily modified to construct an infinite circ-UMFF.
We illustrate the algorithm by extending the border-free UMFF in Lemma \ref{lem_Galois-bf} to a circ-UMFF: 
\begin{examp}
\label{ex-Galois_bf} Given an UMFF $\mathcal{V}$ over $\Sigma = \{a,b\}$ consisting of a subset of border-free Galois words, we enlarge $\mathcal{V}$ to the circ-UMFF $\mathcal{W}$ for words up to length 5; $\mathcal{V}$ and $\mathcal{W}$ are specified below where the Galois words in $\mathcal{W}$ are underlined:
\begin{itemize}
\item $\mathcal{V} = \{a, b, abb, ababb\}$
\item $\mathcal{W} = \{\underline{a}, \underline{b}, \underline{ab}, aab, \underline{abb}, aaab, aabb, \underline{abbb}, aaaab, aaabb,aabab, aabbb, \underline{ababb},\\ 
\underline{abbbb}\}$
\end{itemize}
For instance, string $\s{x} = b b ababb aaabb a  b  aabbb a a$ factors uniquely and maximally:  
\begin{itemize}  
\item over $\mathcal{V}$, $\s{x} =(b) (b) (ababb) (a) (a) (abb) (a)  (b)  (a) (abb) (b) (a) (a)$ 
\item over $\mathcal{W}$, $\s{x} = (b) (b) (ababb) (aaabb) (ab) (aabbb) (a) (a)$
\item but not necessarily uniquely over $\mathcal{G}$, $\s{x} = (b) (b) (ababb aaabb abaabbba a)$
\end{itemize}
\end{examp}

We believe this is the first example of a unique factorization family which is based on more than one type of ordering methodology.


Regarding factoring strings over circ-UMFFs which are derived from mixed methods, such as $\mathcal{W}$ in Example \ref{ex-Galois_bf}, this can be achieved efficiently using the generic algorithmic framework given in \cite{DDIS13}.

\section{Concluding comments}
\label{sec_conclude}


The concept of circ-UMFFs for uniquely factoring strings is a generalization of Lyndon words, which are known to be border-free string conjugates. The literature includes instances  of circ-UMFFs for both regular and indeterminate (degenerate) strings. In this paper we have extended current knowledge on circ-UMFF theory including a further generalization to substring circ-UMFFs. Known instances of circ-UMFFs have been defined using a total order over $\Sigma^*$, such as $V$-order for arbitrary alphabets generating $V$-words, binary $B$-order generating $B$-words, and lex-extension order generating indeterminate Lyndon words.\\ 

We establish here that given any total ordering methodology $\mathcal{T}$ over $\Sigma^*$, and a subset of $\Sigma^*$ consisting of border-free conjugates minimal in $\mathcal{T}$-order, the subset defines a circ-UMFF. An analogous result is established for substring circ-UMFFs. The border-free requirement is illustrated using Galois words by showing that they do not necessarily yield unique maximal string factorization --- Galois words are thus worthy of deeper investigation in this context.\\ 

We have also delved further into the relationship between Lyndon and $V$-words, in particular showing that
there are infinitely more $V$-words than Lyndon words. Other novel concepts are illustrated throughout.

\vskip 1cm

\subsection{Open Problems}
\label{subsec_probs}

\begin{enumerate}
    \item 
Non-unique factorization of a string is not necessarily a bad thing, since having multiple factorizations of a given string $\s{x}$, such as with Example 44 in \cite{DBLP:journals/tcs/DolceRR19}, opens avenues for choice. Accordingly, we propose future UMFF-based research into optimization type problems for string factoring over a fixed alphabet $\Sigma$, such as finding the factorization of $\s{x}$ with the least/greatest number of factors. Related research on controlling Lyndon factors and influencing the run number of the Burrows-Wheeler transform by manipulating the order of $\Sigma$ has been investigated in \cite{DBLP:journals/ipl/ClareD19,DBLP:conf/esa/BentleyGT20,DBLP:journals/corr/abs-2401-16435}.

\item Generalize Algorithm~\ref{alg-UMFF2circUMFF} for enlarging a border-free UMFF to a circ-UMFF from binary to general alphabet. 
Observe that Example \ref{ex_primitive} on a ternary alphabet shows that two conjugates from the same conjugacy class --- namely $abc$ and $cab$ --- can belong to an UMFF $\mathcal{W}$. So this $\mathcal{W}$ cannot be extended to a circ-UMFF, which by Definition~\ref{def-subcirc} must contain exactly one conjugate from each conjugacy class. However, as the example shows, $\mathcal{W}$ is not border-free. 

\item Application of distinct string ordering methods over $\Sigma^*$ and associated circ-UMFFs has been studieded in the context of the Burrows-Wheeler transform and indexing techniques \cite{DS14,DBLP:journals/tcs/DaykinGGLLLP16,DaykinGGLLLMPW19,DBLP:journals/tcs/DaykinMS21}. There is much scope for discovery of new UMFFs and circ-UMFFs, and subsequent exploration of new instances, both combinatorially and algorithmically, over a variety of alphabets. 
 
\item 
The novel entity of circ-UMFFs defined using mixed methods identified in this work is worthy of thorough investigation. For instance, with reference to Example~\ref{ex-Galois_bf}, where $\mathcal{W}$ is the union of Galois and Lyndon words, the following question arises:
given UMFFs $\mathcal{X}$, $\mathcal{Y}$, both over $\Sigma$, and defined using ordering method $\mathcal{\Omega}_{\mathcal{X}}$, $\mathcal{\Omega}_{\mathcal{Y}}$, respectively, where $\mathcal{X}$ generates both bordered and border-free words, then for each primitive bordered word \s{b} in $\mathcal{X}$, choose a border-free conjugate of \s{b} from $\mathcal{Y}$ for the new circ-UMFF. Since in Example~\ref{ex-Galois_bf} $\mathcal{\Omega}_{\mathcal{X}}$ is alternating lexorder and $\mathcal{\Omega}_{\mathcal{Y}}$ is lexorder, can $\mathcal{\Omega}_{\mathcal{Y}}$ always be lexorder --- since every conjugacy class of a primitive word contains a border-free Lyndon word?  


\end{enumerate}

 \subsection{Acknowledgements}

Funding: The second and third authors were funded by the Natural Sciences \& Engineering Research Council of Canada [Grant Numbers: RGPIN-2024-06915 and RGPIN-2024-05921, respectively].

\def\AJC{Australasian J.\ Combinatorics\ }
\def\AWOCA{Australasian Workshop on Combinatorial Algs.}
\def\CPM{Annual Symp.\ Combinatorial Pattern Matching}
\def\COCOON{Annual International Computing \& Combinatorics Conference}
\def\FOCS{IEEE Symp.\ Found.\ Computer Science}
\def\AESA{Annual European Symp.\ on Algs.}
\def\LATA{Internat.\ Conf.\ on Language \& Automata Theory \& Applications}
\def\IWOCA{Internat.\ Workshop on Combinatorial Algs.}
\def\AWOCA{Australasian Workshop on Combinatorial Algs.}
\def\STACS{Internat.\ Symp.\ Theoretical Aspects of Computer Science}
\def\ICALP{Internat.\ Colloq.\ Automata, Languages \& Programming}
\def\IJFCS{Internat.\ J.\ Foundations of Computer Science\ }
\def\ISAAC{Internat.\ Symp.\ Algs.\ \& Computation}
\def\SPIRE{String Processing \& Inform.\ Retrieval Symp.}
\def\SWAT{Scandinavian Workshop on Alg.\ Theory}
\def\PSC{Prague Stringology Conf.}
\def\WALCOM{International Workshop on Algorithms \& Computation}
\def\ALG{Algorithmica\ }
\def\CSUR{ACM Computing Surveys\ }
\def\FI{Fundamenta Informaticae\ }
\def\IPL{Inform.\ Process.\ Lett.\ }
\def\IS{Inform.\ Sciences\ }
\def\JACM{J.\ Assoc.\ Comput.\ Mach.\ }
\def\CACM{Commun.\ Assoc.\ Comput.\ Mach.\ }
\def\MCS{Math.\ in Computer Science\ }
\def\NJC{Nordic J.\ Comput.\ }
\def\SICOMP{SIAM J.\ Computing\ }
\def\SIDMA{SIAM J.\ Discrete Math.\ }
\def\JCB{J.\ Computational Biology\ }
\def\JA{J.\ Algorithms\ }
\def\JCMCC{J.\ Combinatorial Maths.\ \& Combinatorial Comput.\ }
\def\JDA{J.\ Discrete Algorithms\ }
\def\JALC{J.\ Automata, Languages \& Combinatorics\ }
\def\SODA{ACM-SIAM Symp.\ Discrete Algs.\ }
\def\SPE{Software, Practice \& Experience\ }
\def\TCJ{The Computer Journal\ }
\def\TCS{Theoret.\ Comput.\ Sci.\ }

\bibliographystyle{splncs04}
\bibliography{references}
\end{document}